\documentclass[11pt,a4paper,fleqn]{article}
\usepackage[utf8]{inputenc}
\usepackage[english]{babel}
\usepackage{flafter}
\usepackage{amsmath}
\usepackage{amsfonts}
\usepackage{amssymb}
\usepackage{graphicx}
\usepackage{nicefrac}
\usepackage[left=3cm,right=2cm,top=2cm,bottom=2cm]{geometry}

\newcommand{\alphaten}{\boldsymbol{\alpha}}
\newcommand{\Aten}{\boldsymbol{A}}
\newcommand{\AAten}{\mathbb{A}}
\newcommand{\Cten}{\mathbb{C}}

\newcommand{\Dten}{\mathbb{D}}
\newcommand{\eten}{\boldsymbol{\varepsilon}}	% strain tensor
\newcommand{\eelten}{\boldsymbol{\varepsilon}^{\text{el}}}
\newcommand{\eplten}{\boldsymbol{\varepsilon}^{\text{pl}}}
\newcommand{\etendot}{\dot{\boldsymbol{\varepsilon}}}	% strain tensor
\newcommand{\eeltendot}{\dot{\boldsymbol{\varepsilon}}^{\text{el}}}
\newcommand{\epltendot}{\dot{\boldsymbol{\varepsilon}}^{\text{pl}}}
\newcommand{\eqpl}{\varepsilon_{\text{q}}^{\text{pl}}}
\newcommand{\eqplbar}{\bar\varepsilon_{\text{q}}^{\text{pl}}}
\newcommand{\eqplbari}{\bar\varepsilon_{\text{q,i}}^{\text{pl}}}
\newcommand{\eqplbarmax}{\bar\varepsilon_{\text{q,\,max}}^{\text{pl}}}
\newcommand{\eqpldot}{\dot{\varepsilon}_{\text{q}}^{\text{pl}}}
\newcommand{\Hten}{\boldsymbol{H}}
\newcommand{\Iten}{\boldsymbol{I}}
\newcommand{\IIten}{\mathbb{I}}
\newcommand{\IItenT}{\mathbb{I}^T}
\newcommand{\IItensym}{\IIten^{\text{sym}}}
\newcommand{\IItenvol}{\IIten^{\text{vol}}}
\newcommand{\IItendev}{\IIten^{\text{dev}}}
\newcommand{\Jvec}{\boldsymbol{J}}
\newcommand{\mse}{\text{mse}}

\newcommand{\ntenbar}{\bar{\boldsymbol{n}}}
\newcommand{\ntenhat}{\hat{\boldsymbol{n}}}
\newcommand{\Mten}{\boldsymbol{M}}
\newcommand{\Nten}{\boldsymbol{N}}
\newcommand{\nullvec}{\boldsymbol{0}}
\newcommand{\Nullten}{\mathbb{O}}
\newcommand{\pvec}{\boldsymbol{p}}
\newcommand{\plmult}{\lambda}
\newcommand{\Rvec}{\boldsymbol{R}}
\newcommand{\sdev}{\boldsymbol{s}}
\newcommand{\sten}{\boldsymbol{\sigma}}
\newcommand{\stenbar}{\bar{\boldsymbol{\sigma}}}
\newcommand{\stendot}{\dot{\boldsymbol{\sigma}}}

\newcommand{\tr}[1]{\text{tr}\left(#1\right)}

\newcommand{\argmin}[1]{\underset{#1}{\operatorname{argmin}}}
\newcommand{\bsf}[1]{\boldsymbol{\mathsf{#1}}}
\newcommand{\norm}[1]{\left\| #1 \right\|}
\newcommand{\pdiff}[2]{\dfrac{\partial{#1}}{\partial{#2}}}
\newcommand{\pddiff}[2]{\dfrac{\partial^2{#1}}{\partial{#2}^2}}
\newcommand{\pdddiff}[3]{\dfrac{\partial^2{#1}}{\partial{#2}\partial{#3}}}

\author{M. Abendroth\footnote{Institut for Mechanics and Fluid Dynamics, TU Bergakademie Freiberg}, A. Malik, B. Kiefer}
\title{A modified Ehlers model for the description of inelastic behavior of porous structures}

\begin{document}
\maketitle
\begin{abstract}This paper describes a modification of Ehlers' model for the inelastic behavior of granular media. The modified model can be applied for describing the inelastic behavior of porous media. The key feature is a subtle change of the yield potential, which allows the correct orientation of the triangular-shaped yield surface cross sections depending on the hydrostatic stress state. The model is incorporated into a general framework for isotropic plasticity. An elastic predictor/corrector algorithm is employed to solve the constitutive equations. The necessary derivatives for a Newton update are also given in detail. The model is calibrated using stress, and strain data obtained from finite element simulations of a generic highly porous open-cell Wheire-Phelan foam.
\end{abstract}
\section{Introduction}
\begin{figure}[htbp]
\centering
\includegraphics[width=0.6\columnwidth,trim={0 0 0 5},clip]{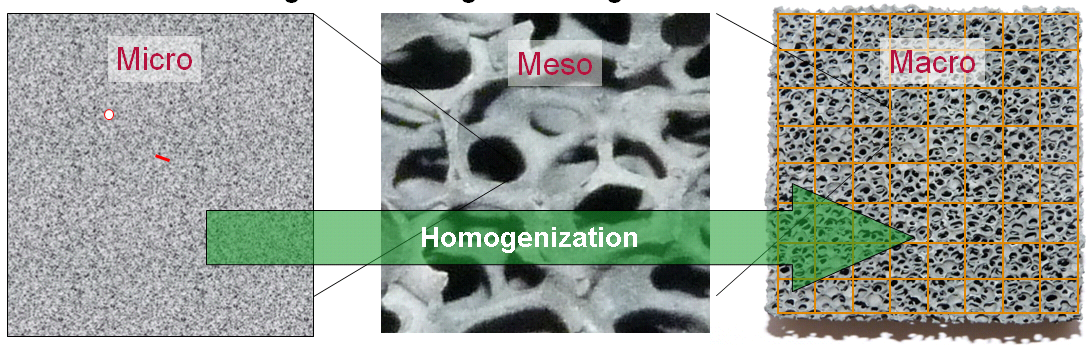}
\caption{\label{fig:HomogenizationScheme}Homogenization scheme for an open cell foam structure.}
\end{figure}

Porous structures or foams appear in many technical applications as well as in nature. Especially for technical applications it is desirable to have computational models for such structures in order to predict their strength, elastic and inelastic behavior. The mechanical properties of porous structures strongly depend on their underlying meso and micro-scale, whereas homogenization schemes as shown in Fig.~\ref{fig:HomogenizationScheme} are applied.

The mechanical behavior on the micro-scale is defined by the properties of the bulk material a foam is made of. Often, the behavior on the micro-scale is known or can be described using established constitutive models. The behavior on the macro scale depends significantly on the topology of the mesostructure, the size and shape of the struts, and their connections \cite{Wang2005, Wang2006}.

Ashby \cite{Ashby2006} investigated the general physical properties of foams and lattices depending on their geometrical structure and topology. He divided between bending and stretching-dominated structures. Critical states of foam structures, such as failure or onset of inelastic deformation can be described by limit surfaces in stress space. A lot of research has been done to describe such surfaces, whereas the majority of publications focus on the transition between elastic and plastic domains (yield surfaces) \cite{GibsonAshby1989} or on surfaces characterizing the onset of material failure (failure surfaces). A comprehensive overview of phenomenological yield and failure surfaces is given in \cite{Altenbach2014}.

Since foams or other porous structures can fail or yield for pure compression as well as for pure tension, their yield or failure surfaces are usually closed and convex surfaces in stress space \cite{GibsonAshby1989}. Since it is difficult and expensive to perform experiments with triaxial stress states or stress states with large hydrostatic components, the experimental data available are limited with respect to the sampled stress space. Jung and Diebels \cite{JungDiebels2018} reviewed contributions with respect to experimental and modeling approaches regarding open cell foams.

One way to generate sufficient data for complete yield and failure surfaces is through direct numerical simulation (DNS) of foam structures. Such techniques have been used in a number of publications e.g.~\cite{Demiray2007, Laroussi2002, Luxner2007, Storm2016, StormAEM2015, Zhang2015}. Depending on the underlying mesostructure, it is observed that yield and failure surfaces of foam structures can depend on the three invariants $I_1$, $J_2$, and $J_3$ of the stress tensor. Especially, the dependence on the third invariant $J_3$ is remarkable. The deviatoric cross sections of the yield surfaces are not necessarily of circular shape but vary from rounded triangular to rounded hexagonal shapes. The triangular shapes can have opposite orientations, depending on the sign of the hydrostatic stress \cite{Abendroth_AEM2017, Demiray2007, FahlbuschGrenestedtBecker2016, FlorenceSab2005, Malik_AEM2022, Wang2006}. The microstructures used in the models for DNSs are often representative volume elements (RVEs) of regular foam structures, like the Kelvin or Wheire-Phelan cell, which also can show a certain anisotropy. But, even yield surfaces for isotropic foam structures \cite{Abendroth_AEM2017} show non-circular deviatoric cross sections.

Constitutive models for foams have been developed by Deshpandy and Fleck \cite{DeshpandeFleck2000}, their model depends only on the first and second invariant of the stress tensor, which results in circular deviatoric cross sections. Öchsner \cite{Oechsner2010} gives a comprehensive introduction to elastic-plastic mechanics of foams, whereas it is also mentioned that the yield surfaces of foams in general can depend on the three stress invariants $I_1$, $J_2$ and $J_3$. Anisotropic failure and yield criteria have been developed by Tsai and Wu \cite{TsaiWu1971}, Barlat \cite{Barlat1991}, Bilkhu \cite{Bilkhu1993} and Nusholtz \cite{Nusholtz1996}. The constitutive behavior of foams becomes more complex if the shape of the yield surface is assumed to depend on internal variables describing the deviatoric and hydrostatic deformation state. Initial and subsequent yield surfaces have been investigated by Demiray \cite{Demiray2007} and Storm \cite{Storm2016}. A data-driven model using neural networks has been developed by Settgast et al.~\cite{Settgast_MOM2019, Settgast_IJP2020, Malik_AEM2022}. Here, the yield surface is approximated by a regression-type neural network. The training data for the network are generated by DNSs of RVEs of foam structures. 

Interestingly, failure and yield surfaces of foams have similar features as those for granular media like sand, rock, or other geomechanic materials.
A very flexible yield or failure criterion for such materials has been proposed by Bigoni and Piccolroaz \cite{Bigoni2004}. It allows various adjustments of the shapes of both hydrostatic and deviatoric yield surface cross sections. As a single surface criterion it cannot provide the necessary shape flip of the deviatoric cross-section along the hydrostatic axis. Bolchoun, Kolupeav and Altenbach \cite{BolchounKolupaevAltenbach2011} provide a comprehensive collection of yield and failure criteria for geomaterials. Especially, their so-called geometric mechanical model (GMM) can describe most of the features of interest. Jung and Diebels proposed among others the Ehlers model \cite{Ehlers1995, EhlersAvci2012} as a potential candidate for yield surfaces of open cell foams. In the following, we will focus on the Ehlers yield surface, but present a modification of the model, which allows us to model the change of the orientation of the triangular cross-section in hydrostatic tension and compression.

The paper is organized as follows: First, the original Ehlers model is recalled, followed by the description of its modification. The essential parameters of the model are discussed and certain restrictions are defined, which ensure the convexity of the yield surface. It follows a section, where a general framework for a constitutive model is presented. This is mainly based on a thermodynamically consistent frame work presented in the book of de Souza Neto \cite{deSouzaNeto2008}. To solve the constitutive equations a general return algorithm is presented, which can be implemented in finite element codes. To apply the modified model to foam structures a parameter identification procedure is necessary. DNSs of a Wheire-Phelan foam are used to gather stress and strain data, which are used to identify parameters of the model and their dependence on the load history. To show the accuracy of the developed model predictions of the model are compared with DNSs of the corresponding foam structures. The article is closed with conclusions and an outlook for improvements and possible extensions of the model. Within an appendix, all necessary derivatives for the implementation of the model as well as for the parameter identification procedure are given.

\section{Modelling}
\subsection{Original Ehlers Model}
The original form of the Ehlers yield surface \cite{Ehlers1995, EhlersAvci2012} has the form
\begin{align}
F &= \sqrt{J_2 \left[1 + \gamma\frac{J_3}{J_2^{\nicefrac{3}{2}}}\right]^m + \frac{1}{2} \alpha I_1^2 + \delta^2 I_1^4} + \beta I_1 + \epsilon I_1^2 - \kappa = 0\,,
\label{eq:EhlersYieldSurface}
\end{align}
which is formulated in the space of the three stress invariants
\begin{align}
I_1 &= \tr{\sten} \,, \\
J_2 &= \frac{1}{2} \sdev : \sdev \,, \\
J_3 &= \det(\sdev)\,,
\end{align}
where $\sten$ denotes the symmetric Cauchy stress tensor, and
\begin{align}
\sdev = \text{dev}(\sten) = \sten - \frac{1}{3} I_1 \Iten
\end{align}
its deviator, with $\Iten$ as the second-order unit tensor. In \eqref{eq:EhlersYieldSurface} $\alpha$, $\beta$, $\delta$ and $\epsilon$ are parameters describing the shape of the meridian cross section, $\gamma$ and $m$ parameters describing the shape the deviatoric cross-section of $F$. Parameter $\kappa$ scales the size of the yield surface. If $\kappa$ depends on the equivalent plastic strain $\eqpl$, it also describes the isotropic hardening behavior. 
The right term in the square brackets of \eqref{eq:EhlersYieldSurface} is related to the Lode angle $\theta$, as
\begin{align}
\frac{J_3}{J_2^{\nicefrac{3}{2}}} &= - \frac{2}{3\sqrt{3}} \sin\left(3\theta\right)\,,
\label{eq:LodeAngle1}
\end{align}
or vice versa, the Lode angle may be expressed using
\begin{align}
\theta &= \frac{1}{3} \sin^{-1} \left( - \frac{3 \sqrt{3}}{2} \frac{J_3}{J_2^{\nicefrac{3}{2}}} \right)\,.
\label{eq:LodeAngle2}
\end{align}
One may note that Eq.~\eqref{eq:LodeAngle2} delivers values for $\theta$ in the range of $[\nicefrac{-\pi}{6},\nicefrac{\pi}{6}]$, whereas for the argument in \eqref{eq:LodeAngle1} any scalar value for $\theta$ can be given. 
\subsection{Modified Ehlers Model}
The modification done to the original Ehlers model realizes a smooth change of parameter $\gamma$ depending on the hydrostatic stress state. In general, it is considered that all parameters can depend on internal variables, especially on $\eqpl$, which allows to model shape changes of the yield surface during a deformation process. The modified version reads as
\begin{align}
F &= \sqrt{J_2 \left[1 + \gamma \, A \, C\right]^m + \frac{1}{2} \alpha I_1^2 + \delta^2 I_1^4} + \beta I_1 + \epsilon I_1^2 - \kappa = 0\,,
\label{eq:ModEhlersYieldSurface}
\end{align}
with the additional term
\begin{align}
A &= \frac{\tr{\Nten}}{\sqrt{3}} = \sin \left( \tan^{-1} \left( \frac{I_1}{\sqrt{6\,J_2}} \right) \right) \,,
\label{eq:termA}
\end{align}
whereas $\Nten$ represents the normalized stress or stress direction
\begin{align}
\Nten &= \frac{\sten}{\norm{\sten}} \quad \text{with} \quad \norm{\sten}=\sqrt{\sten : \sten}\,.
\end{align}
For convenience the term
\begin{align}
C &= \frac{J_3}{J_2^{\nicefrac{3}{2}}} = \frac{2}{3\sqrt{3}} \sin\left(3\theta\right)
\end{align}
is introduced, which simplifies the notation later on.
The term $A$ changes smoothly in the range $[-1,1]$ depending on the hydrostatic stress. The sign change of $A$ realizes the flip of the triangular-shaped deviatoric cross-section of the yield surface with respect to the state of the hydrostatic stress, as it is observed for yield surfaces of foam structures \cite{Malik_AEM2022}. $F$ can be also understood as a yield and/or flow potential for a constitutive model.
The shape of the deviatoric cross section $(I_1=\text{const.})$ can be expressed as
\begin{align}
F^{\text{dev}} &= \sqrt{J_2 \left[1 + \gamma \, A \, C \right]^m}\,.
\end{align}
If Eq.~\eqref{eq:ModEhlersYieldSurface} is solved for $J_2$, an expression for the shape in the hydrostatic plane can be derived, which reads as
\begin{align}
F^{\text{hyd}} &= \sqrt{
\frac{\left( \epsilon^2 - \delta^2 \right) I_1^4 + 2 \, \beta \, \epsilon \, I_1^3 + \left( \beta^2 - \frac{1}{2} \alpha - 2 \, \epsilon \, \kappa \right) I_1^2 - 2 \, \beta \, \kappa \, I_1 + \kappa^2}
{\left[1 + \gamma \, A \, C \right]^m}}
\label{eq:Fhyd}
\end{align}
\begin{figure}[htbp]
\centering
\includegraphics[width=0.48\textwidth]{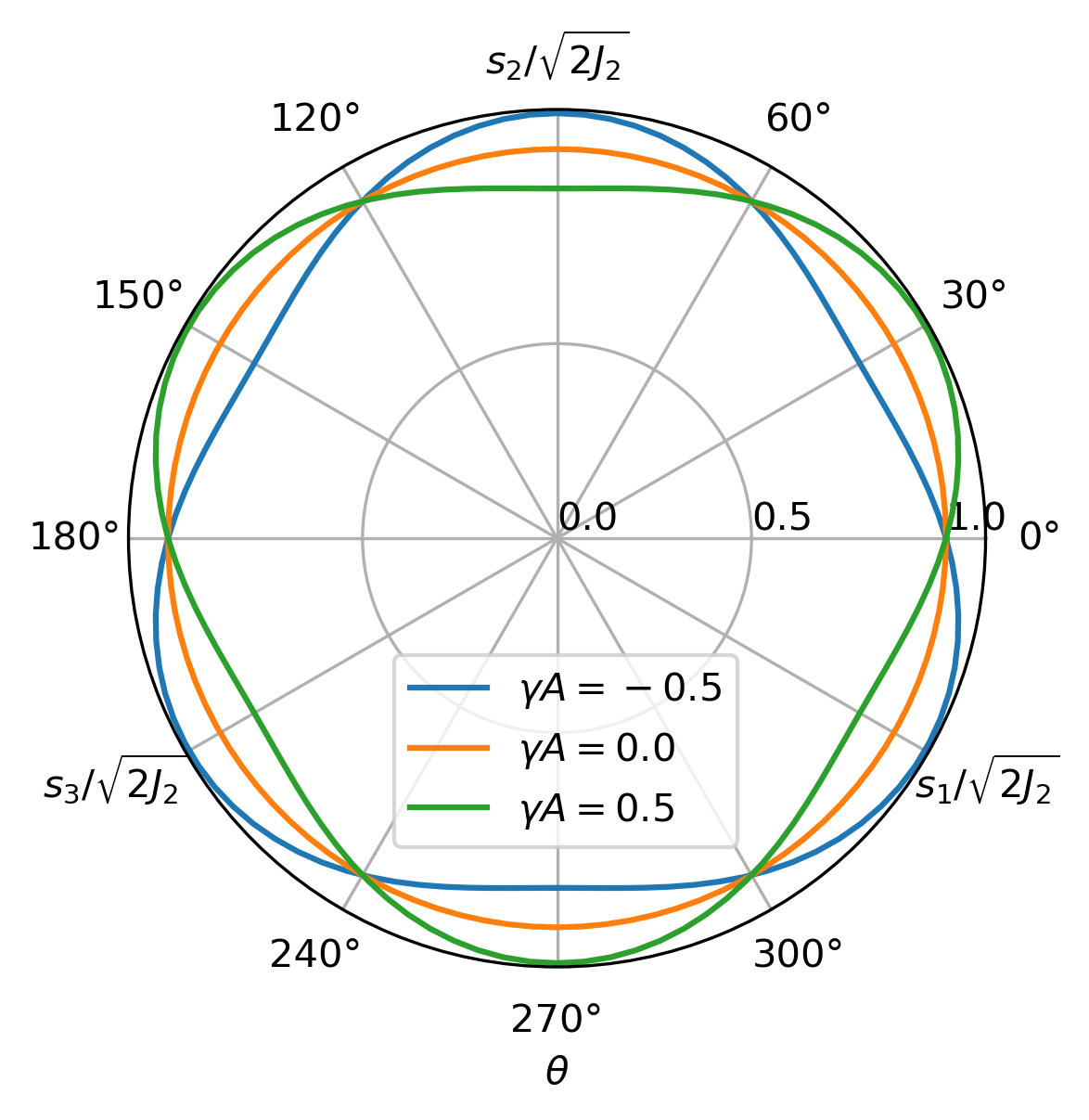}\hfill
\includegraphics[width=0.48\textwidth]{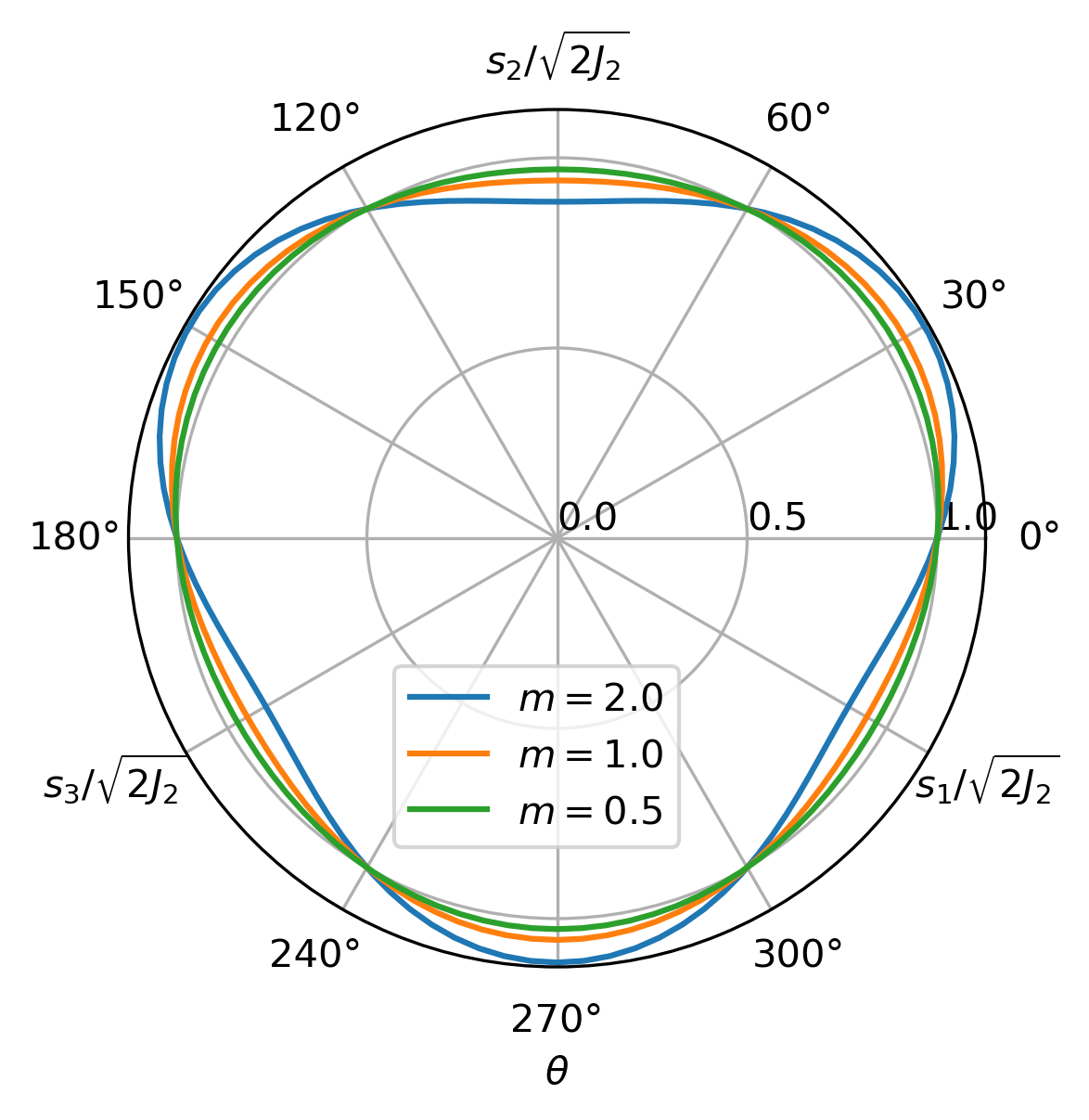}
\caption{\label{fig:Fdev}Deviatoric cross sections of the yield surface for the modified model. Left) For a term $A=[-1,\,0,\,1]$ and parameters $\gamma=0.5$ and $m=1$. Right) For the parameters $\gamma A=0.3$ and a varied parameter $m$.}
\end{figure}

Fig.~\ref{fig:Fdev} shows the shape of the deviatoric cross-section of the yield surface projected onto the deviatoric plane in principal stress space for varied parameters $\gamma$ and $m$. If $\gamma \ne 0$ the deviatoric cross section shows a triangular shape, which becomes more pronounced if the absolute value of $\gamma$ increases. The sign of $\gamma A$ controls the orientation of the shape, having a sharp tip either at an angle $\theta$ of 30° or 90°. Parameter $m$ controls the sharpness of the tips of the yield surfaces' cross-section.
\begin{figure}[htbp]
\centering
\includegraphics[width=0.75\textwidth]{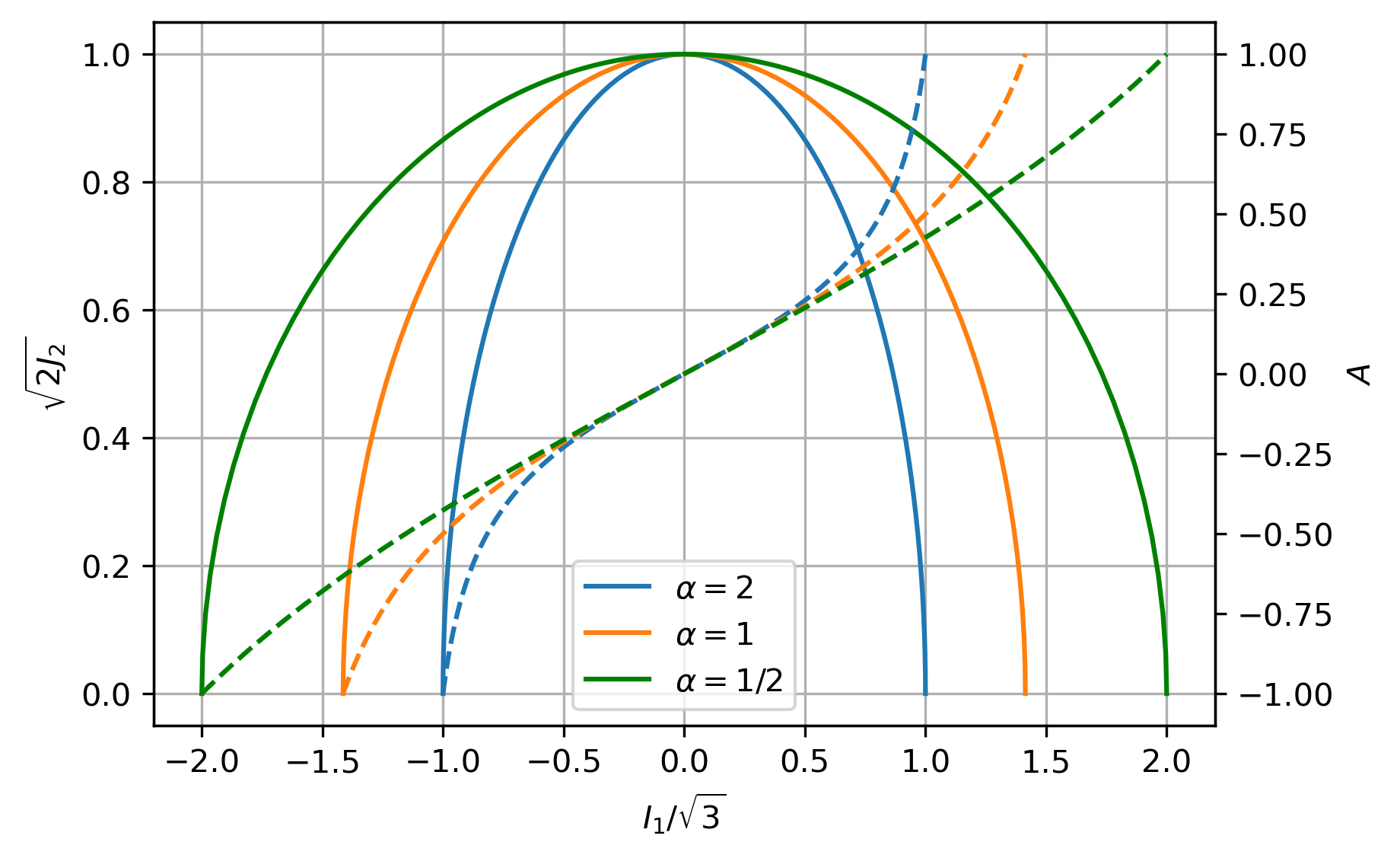}
\caption{\label{fig:FmerA}Meridian cross sections of the yield surface for the modified model for varied parameter $\alpha$. The dashed lines show how the term $A$ for the modified model depending on $I_1$.}
\end{figure}

Fig.~\ref{fig:FmerA} shows the meridian cross-section of the yield surface from the modified model for varied values of parameter $\alpha$. For smaller values of $\alpha$ the yield surface extends symmetrically in the hydrostatic stress direction forming an elongated ellipsoid. The dashed lines show the value of the term $A=\frac{\text{tr}(\Nten)}{\sqrt{3}}$ in Eq.~\eqref{eq:ModEhlersYieldSurface} depending on the hydrostatic stress state expressed by $I_1$. As more elongated the yield surface becomes, as less curved that dependence appears. An important feature of $A$ is, that it a priori changes from $-1$ to $1$ within the bounds of $I_1$ given by the shape of the yield surface.

\begin{figure}[htbp]
\centering
\includegraphics[width=0.45\textwidth]{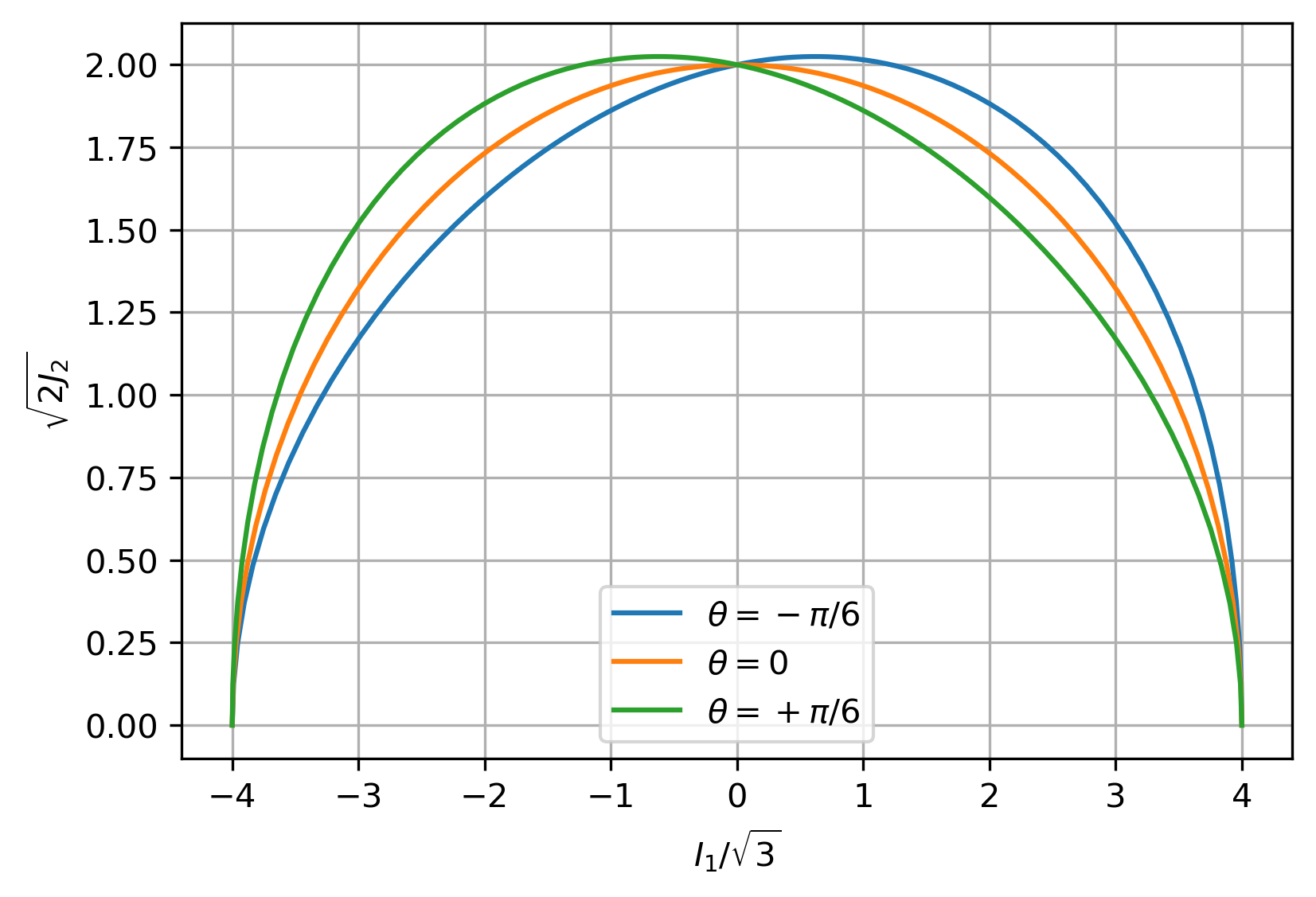} \hfill
\includegraphics[width=0.45\textwidth]{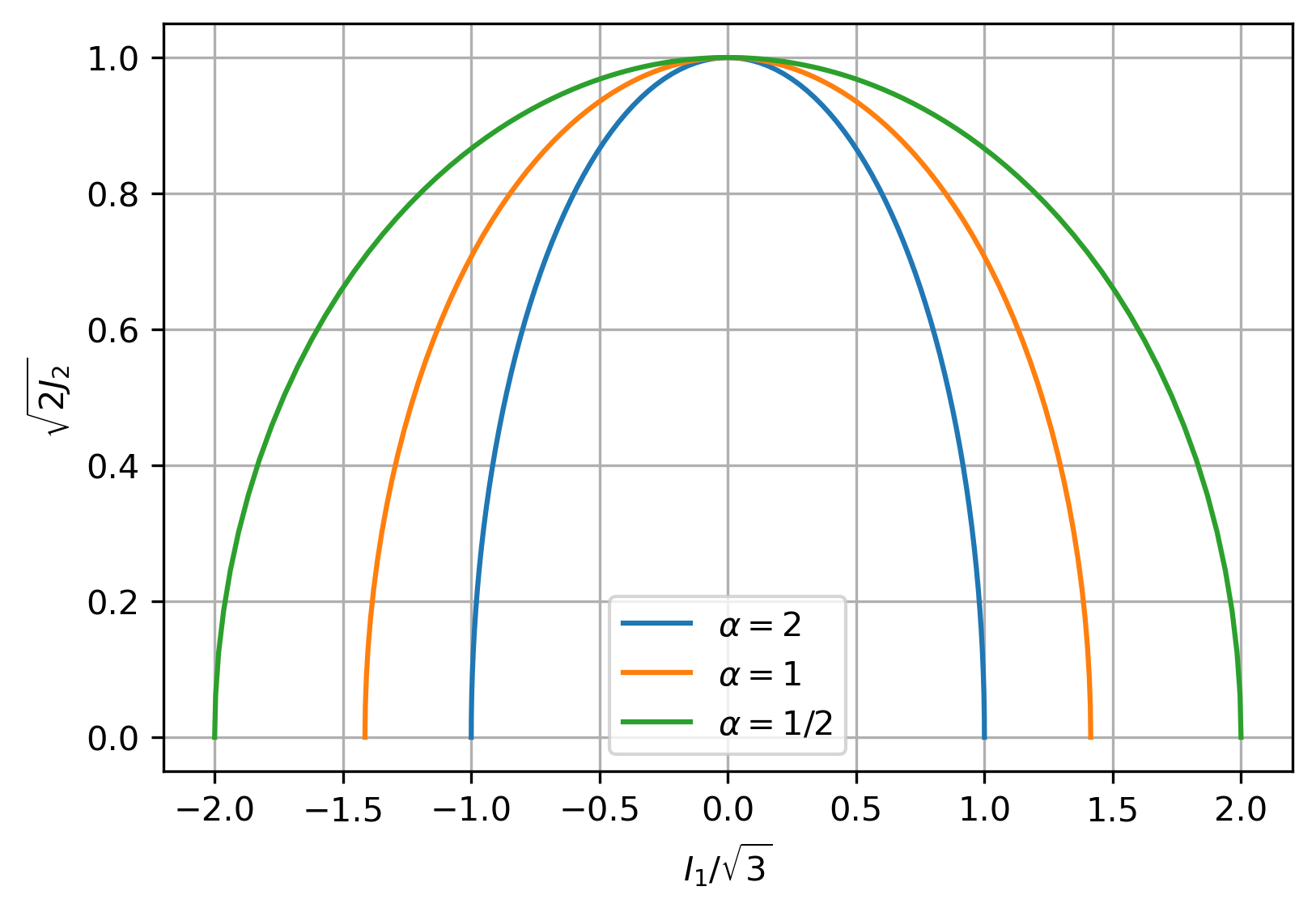} \\
\includegraphics[width=0.45\textwidth]{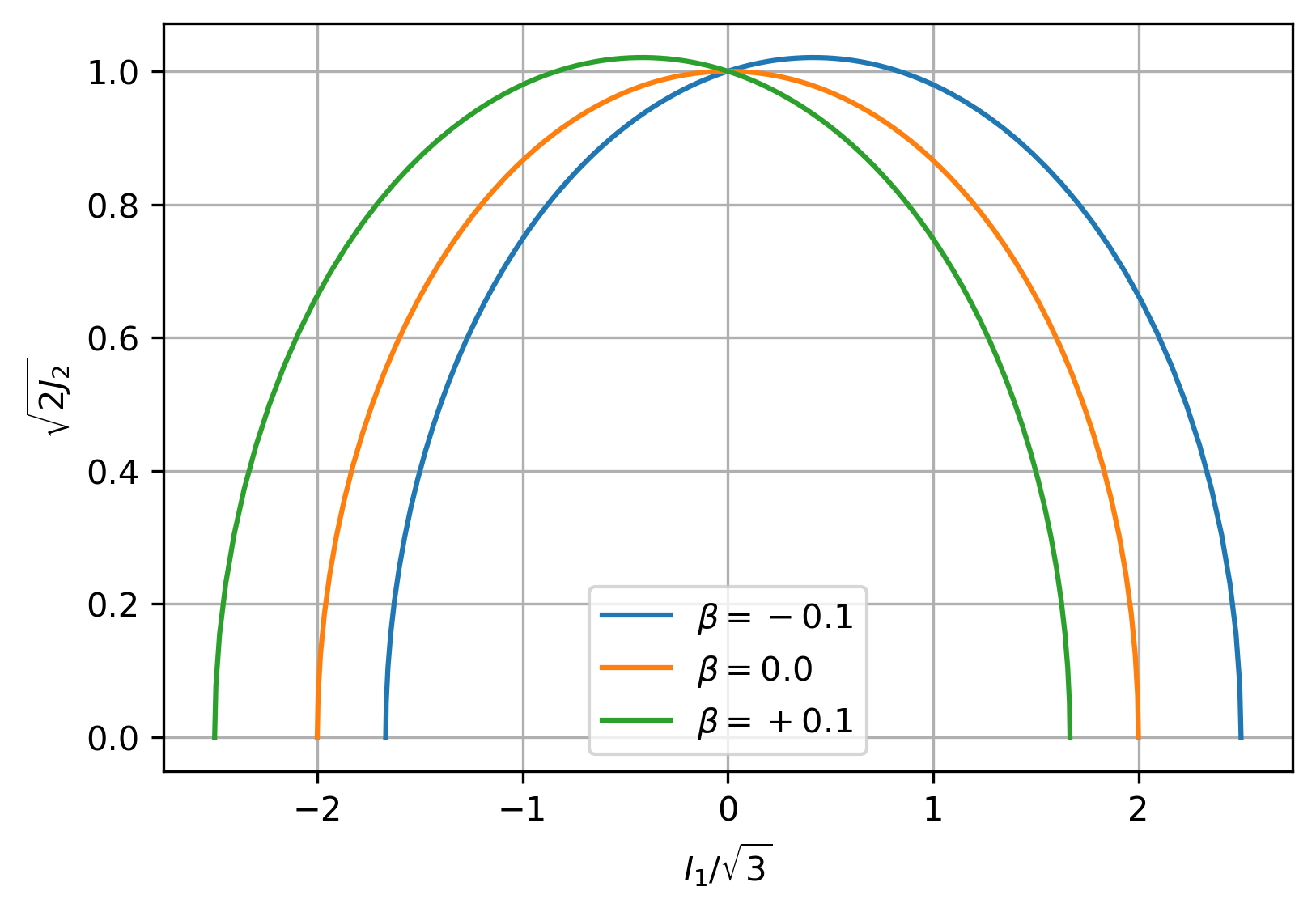} \hfill
\includegraphics[width=0.45\textwidth]{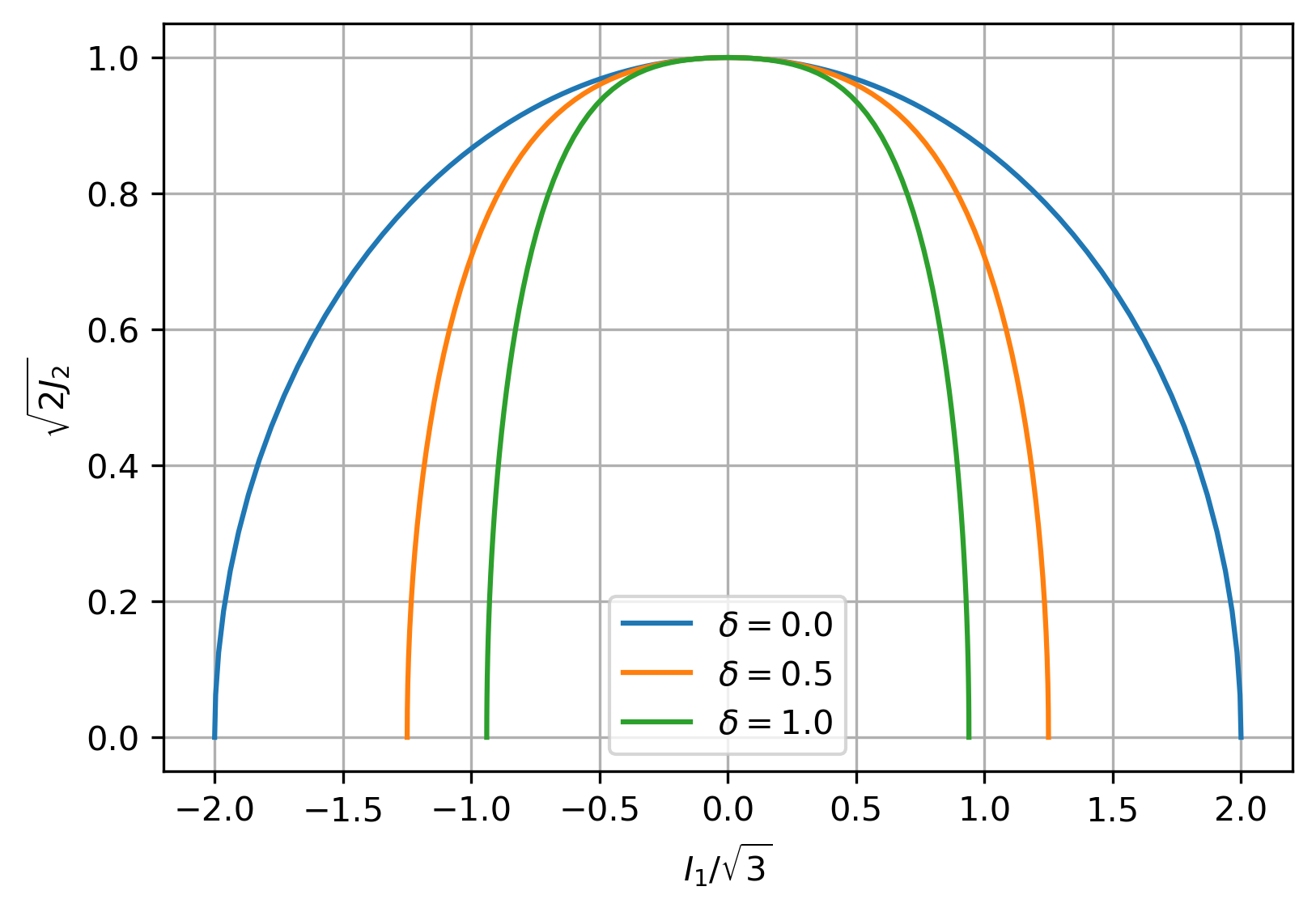} \\
\includegraphics[width=0.45\textwidth]{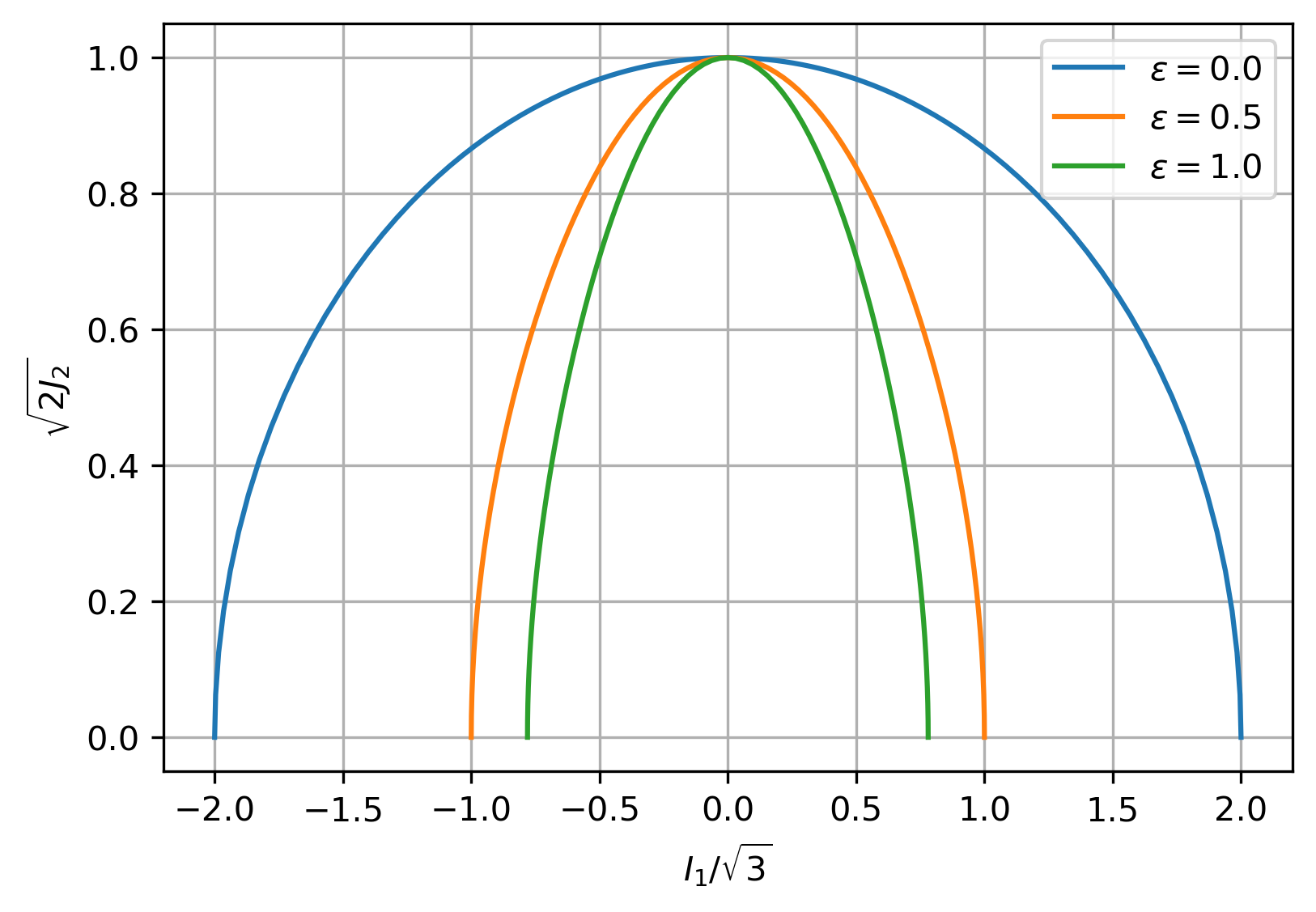} \hfill
\includegraphics[width=0.45\textwidth]{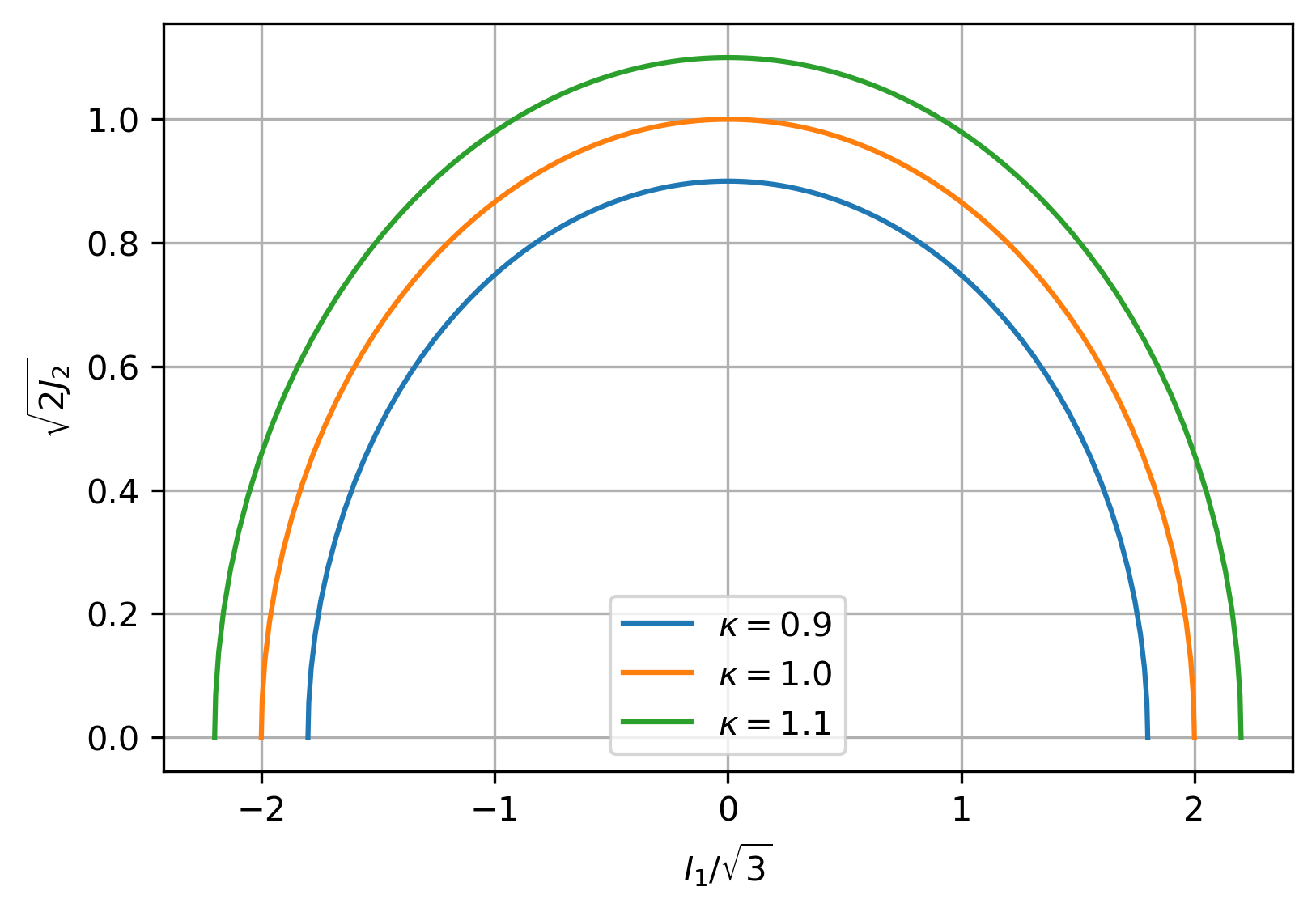} \\
\caption{\label{fig:Fmer}Meridian cross sections of the yield surface for the modified model for varied parameters $\theta$, $\alpha$, $\beta$, $\delta$, $\epsilon$, and $\kappa$.}
\end{figure}

In Fig.~\ref{fig:Fmer} the influences of the parameters $\theta$, $\alpha$, $\beta$, $\delta$, $\epsilon$, and $\kappa$ on the meridian cross-section shape of the yield surface are displayed. The default parameters are set as $\theta=0$, $\alpha=\nicefrac{1}{2}$, $\beta=0$, $\gamma=1$, $\delta=0$\,MPa$^{-1}$, $\epsilon=0$\,MPa$^{-1}$, $\kappa=1$\,MPa, and $m=1$.
For a negative value of $\theta$ the meridian cross section is slightly thinner for negative hydrostatic stresses. Vice versa, for positive $\theta$ values the cross section is thinner for positive hydrostatic stresses. The meridian cross-section is symmetric if $\theta=0$. Decreasing values of $\alpha$ lead to an elongated yield surface along the hydrostatic axis. The parameter $\beta$ shifts the cross-section along the hydrostatic axis. The parameters $\delta$ and $\epsilon$, both change the curvature of the meridian cross section. If both parameters are zero the cross section has an elliptical form. Finally, parameter $\kappa$ scales the whole yield surface in a self-similar manner. For the use as a yield potential, $F$ is required to be strictly convex, which requires certain restrictions on the parameters \cite{Ehlers1995}. The conditions for a convex deviatoric cross-section are
\begin{align}
\gamma &\le \frac{\sqrt{27}}{9\,m-2} \quad \text{or} \quad m \le \frac{\frac{\sqrt{27}}{\gamma}+2}{9} \quad \text{and} \quad \gamma \le \frac{\sqrt{27}}{2}\,.
\label{eq:ConvexityCondFdev}
\end{align}
A sufficient condition for the convexity in the hydrostatic plane is derived from Eq.~\eqref{eq:Fhyd}. Convexity in the hydrostatic plane is ensured, if the second derivative of \eqref{eq:Fhyd} with respect to $I_1$ is smaller or equal to zero.
\begin{align}
\pddiff{F_{\text{hyd}}}{I_1} &\le 0
\label{eq:ConvexityCondFhyd}
\end{align}
This condition is a priori fulfilled if the parameters $\alpha \ge 0$, $\delta \ge 0$, $\epsilon \ge 0$ and $\kappa \ge 0$.

\begin{figure}[htbp]
\centering
\includegraphics[width=0.49\textwidth]{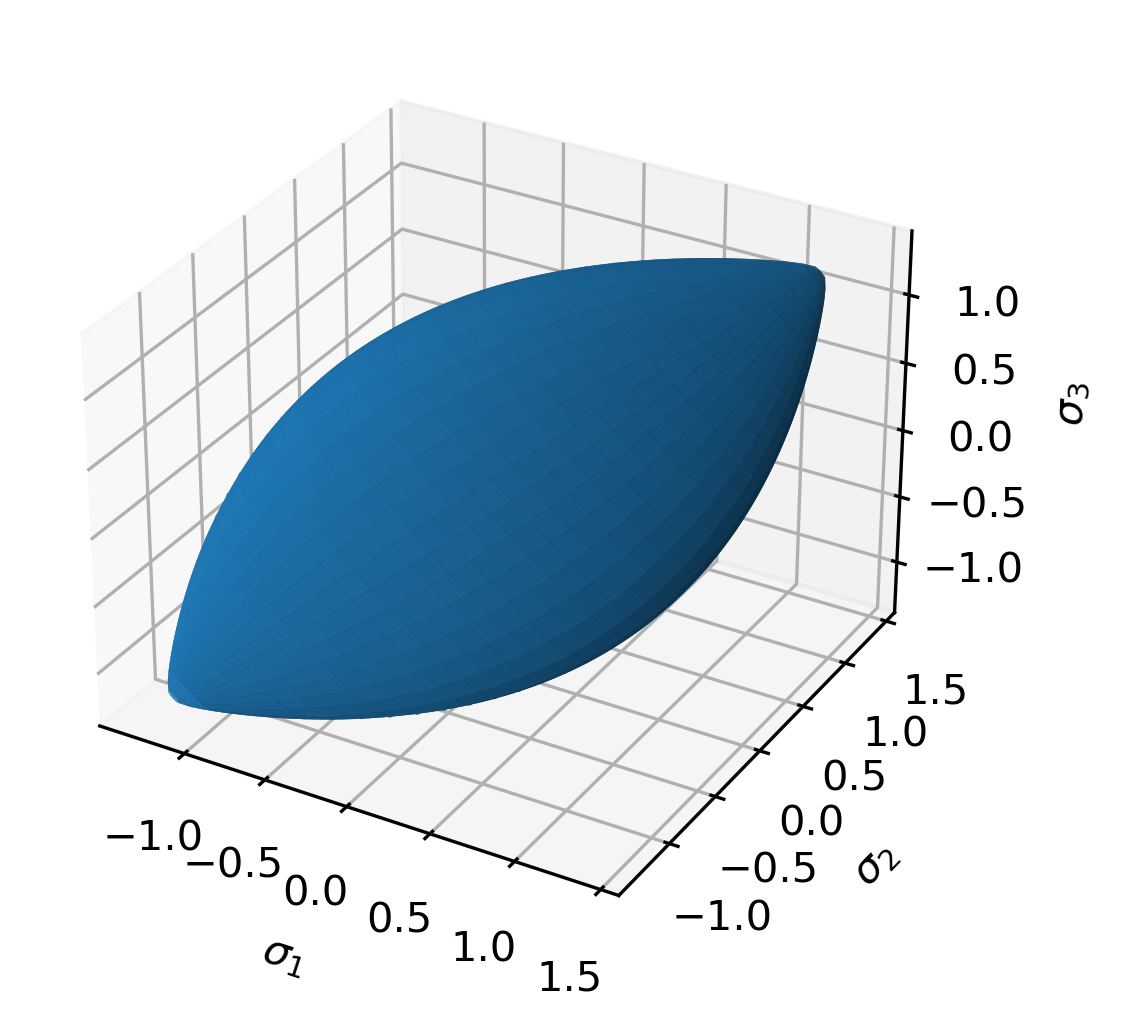} \hfill
\includegraphics[width=0.49\textwidth]{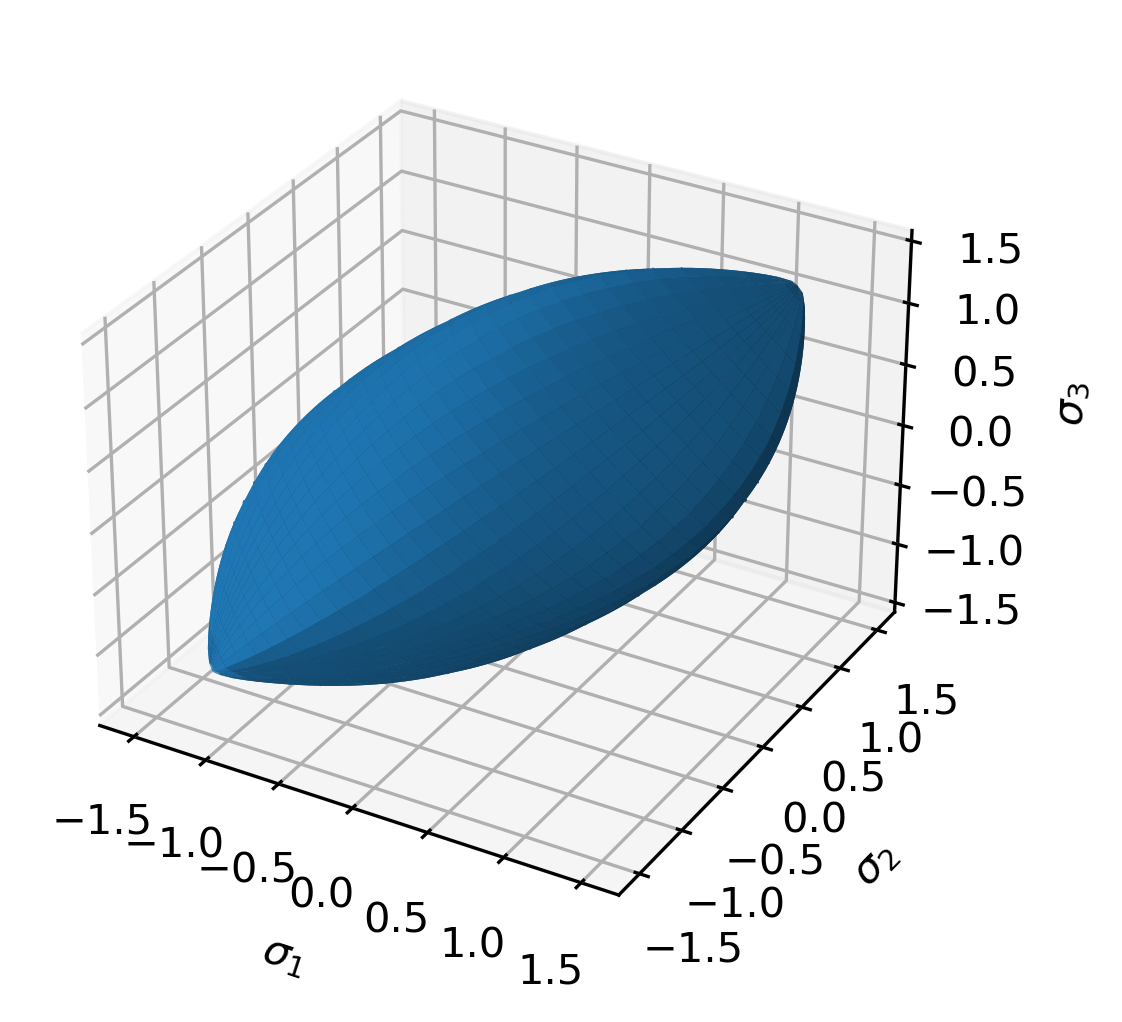} \\
\caption{\label{fig:F_org_mod}Yield surfaces of the original (left) and modified (right) model in principal stress space for the same set of parameters. The yield surface of the original model has a rounded triangular cross-section, which does not change it's orientation depending on the hydrostatic stress state. The modified yield surface has a circular cross-section for zero hydrostatic stress and triangular-shaped cross-sections with opposite orientations for positive and negative hydrostatic stresses.}
\end{figure}

The complete yield surface can be projected into the principal stress space. Fig.~\ref{fig:F_org_mod} shows the original and the modified yield surface for the same set of parameters $\alpha=10^{-8}$, $\beta=0$, $\gamma=2.273$, $\delta=0.0031$\,MPa$^{-1}$, $\epsilon=0.0517$\,MPa$^{-1}$, $\kappa=0.769$\,MPa, and $m=0.389$. These parameters have been identified for an artificial foam structure \cite{Malik_AEM2022}.

\section{A Constitutive Framework for the Inelastic Behavior of Porous Structures}
\subsection{Constitutive Equations}
For the derivation of the constitutive equations within a thermodynamically consistent framework we follow the book of de Souza Neto \cite{deSouzaNeto2008}.
We consider a small deformation setting. The strain rate tensor is additively decomposed into an elastic and plastic part.
\begin{align}
\etendot &= \eeltendot + \epltendot
\label{eq:AdditiveSplit}
\end{align}
A free energy function depending on the elastic strain tensor and a set of internal variables is defined, which is also split into an elastic and a plastic part.
\begin{align}
\psi &= \psi \left( \eelten,\alphaten\right)
= \psi^{\text{el}}\left( \eelten\right)+\psi^{\text{pl}}\left(\alphaten\right) 
= \psi^{\text{el}}\left( \eten - \eplten \right)+\psi^{\text{pl}}\left(\alphaten\right)
\label{eq:FreeEnergy}
\end{align}
A set of internal variables is denoted by $\alphaten$. The corresponding Clausius-Duheme inequality reads
\begin{align}
\left(\sten - \bar\rho \pdiff{\psi^{\text{el}}}{\eelten}\right) : \eeltendot + \sten : \epltendot - \Aten \cdot \dot\alphaten \ge 0\,.
\label{eq:ClausiusDuhem}
\end{align}
Eq.~\eqref{eq:ClausiusDuhem} implies a general elastic law
\begin{align}
\sten &= \bar\rho \pdiff{\psi^{\text{el}}}{\eelten}\,,
\end{align}
and a general hardening thermodynamical force
\begin{align}
\Aten &= \bar\rho \pdiff{\psi^{\text{pl}}}{\alphaten}\,.
\end{align}
The plastic dissipation function is derived from \eqref{eq:ClausiusDuhem} and reads
\begin{align}
\Upsilon^{\text{pl}} &= \sten:\epltendot - \Aten \cdot \dot\alphaten \ge 0\,.
\label{eq:PlasDissipation}
\end{align}
The elastic contribution to the free energy \eqref{eq:FreeEnergy} is given by
\begin{align}
\bar\rho \psi^{\text{el}} &= \frac{1}{2} \eelten : \Cten : \eelten\,,
\end{align}
where $\Cten$ denotes the isotropic forth order stiffness tensor
\begin{align}
\Cten &= 2G\,\IItendev + 3K\,\IItenvol\,,
\end{align}
with the shear and bulk moduli $G$ and $K$, respectively. The fourth order projection tensors $\IItendev$ and $\IItenvol$ are used for the composition of the stiffness tensor. Using the elastic modulus $E$ and Poisson's ratio $\nu$ shear and bulk modulus can be expressed as
\begin{align}
G &= \frac{E}{2(1+\nu)} \quad \text{and} \quad K = \frac{E}{3(1-2\nu)} \,.
\end{align}
The general isotropic elastic law in rate form reads
\begin{align}
\stendot=\Cten : \eeltendot\,,
\end{align}
which is also known as Hooke's law.
The plastic contribution to the free energy \eqref{eq:FreeEnergy} is given by
\begin{align}
\bar\rho \psi^{\text{pl}} &= F\left(\sten,\alphaten\right)\,, 
\end{align}
where $F\left(\sten,\alphaten\right)$ is the yield potential.
The plastic flow rule and the generalized hardening law are given by
\begin{align}
\epltendot &= \dot\plmult \Nten\left(\sten,\Aten\right) \label{eq:PlasFlowRule} \\
\dot\alphaten &= \dot\plmult \Hten\left(\sten,\Aten\right) \label{eq:GenHardLaw}
\end{align}
The tensor $\Nten$ denotes the plastic flow direction, which is derived from a plastic flow potential $G$ as
\begin{align}
\Nten &= \pdiff{G}{\sten}\,.
\end{align}
In case $G=F$ the plastic flow is called associated. If the generalized hardening modulus $\Hten$ is derived from the yield potential $F$
\begin{align}
\Hten &= -\pdiff{F}{\Aten}\,,
\label{eq:GeneralizedHardeningModulus}
\end{align}
Eq.~\eqref{eq:GeneralizedHardeningModulus} is called an associative hardening law.
The constitutive equations are completed by the loading and unloading (Karush-Kuhn-Tucker) conditions
\begin{align}
F &\le 0, \quad \dot\plmult \ge 0, \quad \dot\plmult F = 0\,.
\label{eq:KKT-conditions}
\end{align}

\subsection{General Return Algorithm}
The general return algorithm considered here is adopted from de Souza Neto \cite{deSouzaNeto2008}. First, the general outline of the algorithm is given, followed by a detailed adaption to the specific problem considered in this paper. A time interval $[t_n,t_{n+1}]$ is considered, where it is assumed that at $t_n$ all quantities $\sten_n$, $\eten_n$, $\alphaten_n$ are known. For a deformation driven process, a strain increment $\Delta \eten=\eten_{n+1}-\eten_n$ is given.  To compute the values $\sten_{n+1}$, $\eten_{n+1}$, $\alphaten_{n+1}$ at the end of the time increment we have to solve the following system of equations
\begin{align}
\eelten_{n+1} &= \eelten_n + \Delta \eten - \Delta \plmult \Nten(\sten_{n+1},\Aten_{n+1}) \label{eq:GeneralReturn1}\\
\alphaten_{n+1} &= \alphaten_{n} + \Delta \plmult \Hten(\sten_{n+1},\Aten_{n+1})
\label{eq:GeneralReturn2}
\end{align}
for the unknowns $\eelten_{n+1}$, $\alphaten_{n+1}$, and $\Delta \plmult$, subject to the constraints (Karush-Kuhn-Tucker conditions)
\begin{align}
\Delta \plmult \ge 0, \quad F(\sten_{n+1},\Aten_{n+1}) \le 0, \quad \Delta \plmult \, F(\sten_{n+1},\Aten_{n+1}) = 0\,,
\label{eq:GeneralReturnKKT}
\end{align}
where
\begin{align}
\sten_{n+1} &= \bar\rho \left.\pdiff{\psi}{\eelten}\right|_{n+1}, \quad \Aten_{n+1}=\bar\rho \left.\pdiff{\psi}{\alphaten}\right|_{n+1}\,.
\label{eq:PotentialEquations}
\end{align}
The increment $\Delta \plmult$ is called the incremental plastic multiplier. Once the solution $\eelten_{n+1}$ has been obtained, the plastic strain at $t_{n+1}$ can be calculated as
\begin{align}
\eplten_{n+1} &= \eplten_n + \Delta \eten - \Delta \eelten\,.
\end{align}
The KKT-conditions \eqref{eq:GeneralReturnKKT} allow only two distinct mutually exclusive cases. If $\Delta \plmult=0$ there is no plastic flow nor any evolution of internal variables within the interval $[t_n,t_{n+1}]$, which means that the step is purely elastic. Then the solution is simply given by
\begin{align}
\eelten_{n+1} &= \eelten_n + \Delta \eten\,,\label{eq:GeneralReturnElasticSolution_1}\\
\alphaten_{n+1} &= \alphaten_{n}\,.
\label{eq:GeneralReturnElasticSolution_2}
\end{align}
In case that the plastic multiplier $\Delta \plmult > 0$ the solution for $\eelten_{n+1}$, $\alphaten_{n+1}$, and $\Delta \plmult$ must satisfy the equations \eqref{eq:GeneralReturn1} and \eqref{eq:GeneralReturn2} in combination with the constraints \eqref{eq:GeneralReturnKKT}$_2$ and \eqref{eq:GeneralReturnKKT}$_3$, which results in the system of equations
\begin{align}
&\eelten_{n+1} = \eelten_n + \Delta \eten - \Delta \plmult \Nten(\sten_{n+1},\Aten_{n+1}) \nonumber \\
&\alphaten_{n+1} = \alphaten_{n} + \Delta \plmult \Hten(\sten_{n+1},\Aten_{n+1}) \label{eq:GeneralReturnPlasticSolution} \\
&F(\sten_{n+1},\Aten_{n+1}) = 0\,.
\nonumber
\end{align}
To solve the system \eqref{eq:GeneralReturnPlasticSolution} a fully implicit elastic predictor/corrector return mapping algorithm is employed. The elastic trial state is defined by the state variables at $t_n$ and a given strain increment $\Delta \eten$.
\begin{align}
&\eten_{n+1}^{\text{el trial}} = \eelten_n + \Delta \eten, \quad \alphaten_{n+1}^{\text{trial}} = \alphaten_n \label{eq:TrialState1}\\
&\sten_{n+1}^{\text{trial}} = \bar\rho \left.\pdiff{\psi}{\eelten}\right|_{n+1}^{\text{trial}}, \quad \Aten_{n+1}^{\text{trial}}=\bar\rho \left.\pdiff{\psi}{\alphaten}\right|_{n+1}^{\text{trial}}
\label{eq:TrialState2}
\end{align} 
If $F(\sten_{n+1}^{\text{trial}},\Aten_{n+1}^{\text{trial}}) \le 0$ then the solution is the trial state $( \cdot )_{n+1} = ( \cdot )_{n+1}^{\text{trial}}$, corresponding to an elastic step with update relations \eqref{eq:GeneralReturnElasticSolution_1} and \eqref{eq:GeneralReturnElasticSolution_2}.
If $F(\sten_{n+1}^{\text{trial}},\Aten_{n+1}^{\text{trial}}) > 0$ a plastic step is considered, and by putting the trial state from \eqref{eq:TrialState1} into Eqs.~\eqref{eq:GeneralReturnPlasticSolution} the following system of equations can be set up
\begin{align}
&\left\{
\begin{array}{c}
\eelten_{n+1} - \eten_{n+1}^{\text{el trial}} + \Delta \plmult \Nten_{n+1}\\[1ex]
\alphaten_{n+1} - \alphaten_{n+1}^{\text{trial}} - \Delta \plmult \Hten_{n+1}\\[1ex]
\Phi(\sten_{n+1},\Aten_{n+1})
\end{array}
\right\}
=
\left\{
\begin{array}{c}
\nullvec\\[1ex]
\nullvec\\[1ex]
0
\end{array}
\right\}\,,
\label{eq:ReturnMappingEqs}
\end{align}
which has to be solved for $\eelten_{n+1}$, $\alphaten_{n+1}$, and $\Delta \plmult$ with
\begin{align}
\sten_{n+1} &= \bar\rho \left.\pdiff{\psi}{\eelten}\right|_{n+1}, \quad \alphaten_{n+1}=\bar\rho \left.\pdiff{\psi}{\Aten}\right|_{n+1}\,.
\end{align}
Linearization of Eqs.~\eqref{eq:ReturnMappingEqs} yields
\begin{align}
&\left\{
\begin{array}{c}
\text{d}\eelten + \Delta \plmult \pdiff{\Nten}{\sten} : \text{d}\sten + \Delta \plmult \pdiff{\Nten}{\Aten} \cdot \text{d}\Aten + \text{d} \Delta \plmult \Nten\\[2ex]
\text{d}\alphaten - \Delta \plmult \pdiff{\Hten}{\sten} \cdot \text{d}\sten - \Delta \plmult \pdiff{\Hten}{\Aten} \cdot \text{d}\Aten - \text{d} \Delta \plmult \Hten\\[2ex]
\pdiff{F}{\sten} : \text{d}\sten + \pdiff{F}{\Aten} \cdot \text{d}\Aten
\end{array}
\right\}
=
\left\{
\begin{array}{c}
\text{d}\eten^{\text{el trial}}\\[3ex]
\nullvec\\[3ex]
0
\end{array}
\right\}\,,
\label{eq:LinReturnMapping}
\end{align}
From the potential equations the differentials $\text{d}\sten$ and $\text{d}\Aten$ can be derived (see \cite{deSouzaNeto2008} p.~239).
\begin{align}
\text{d}\sten &= \Cten : \text{d}\eelten + \bsf{E} \cdot \text{d}\alphaten  \nonumber \\
\text{d}\Aten &= \bsf{F} \cdot  \text{d}\eelten + \bsf{G} \cdot \text{d}\alphaten \label{eq:DifferentialsdSdA}
\end{align}
The linear operators $\Cten$, $\bsf{E}$, $\bsf{F}$ and $\bsf{G}$ are defined as
\begin{align}
\Cten &= \bar\rho \pddiff{\psi}{\eelten}, \quad
\bsf{E} = \bar\rho \pdddiff{\psi}{\eelten}{\alphaten}, \quad
\bsf{F} = \bar\rho \pdddiff{\psi}{\alphaten}{\eelten}, \quad
\bsf{G} = \bar\rho \pddiff{\psi}{\alphaten}\,.
\label{eq:LinOperatorsCEFG}
\end{align}
Inversion of \eqref{eq:DifferentialsdSdA} yields the exressions
\begin{align}
\text{d}\eelten &= \Dten : \text{d}\sten + \bsf{B} \cdot \text{d}\Aten  \nonumber \\
\text{d}\alphaten &= \bsf{A} \cdot  \text{d}\sten + \bsf{J} \cdot \text{d}\Aten\,.
\label{eq:Differentialsdeda1}
\end{align}
Since the potential $\psi$ is split additively into an elastic and a plastic part \eqref{eq:FreeEnergy} the tangent moduli $\bsf{E}$ and $\bsf{F}$ as well as $\bsf{A}$ and $\bsf{B}$ vanish and the Eqs.~\eqref{eq:DifferentialsdSdA} can be inverted into
\begin{align}
\text{d}\eelten &= \Dten : \text{d}\sten \nonumber \\
\text{d}\alphaten &= \bsf{J} \cdot \text{d}\Aten\,, \label{eq:Differentialsdeda2}
\end{align}
where $\Dten=\Cten^{-1}$ and $\bsf{J}=\bsf{G}^{-1}$.
Substituting \eqref{eq:DifferentialsdSdA} and \eqref{eq:Differentialsdeda2} into \eqref{eq:LinReturnMapping} yields the symbolic matrix representation
\begin{align}
&
\begin{bmatrix}
\Dten + \Delta \plmult \pdiff{\Nten}{\sten} & \bsf{B} + \Delta \plmult \pdiff{\Nten}{\Aten} & \Nten \\[2ex]
\bsf{A} - \Delta \plmult \pdiff{\Hten}{\sten} & \bsf{J} - \Delta \plmult \pdiff{\Hten}{\Aten} & -\Hten \\[2ex]
\pdiff{F}{\sten} & \pdiff{F}{\Aten} & 0
\end{bmatrix}
\begin{bmatrix}
\text{d}\sten \\[2ex]
\text{d}\Aten \\[2ex]
\text{d}\Delta\plmult
\end{bmatrix}
=
\begin{bmatrix}
\text{d}\eten^{\text{el trial}} \\[2ex]
\nullvec \\[2ex]
0
\end{bmatrix}\,.
\label{eq:LinReturnMappingMatrix}
\end{align}
Eq.~\eqref{eq:LinReturnMappingMatrix} represents the general case, where $\Nten$ and $\Hten$ can depend on $\sten$ and $\Aten$. 

\subsection{Application to the Modified Ehlers Model}
In the special case considered here, the set of internal variables contains only the equivalent plastic strain $\alphaten=\{\eqpl\}$ and we define that $\dot\plmult=\eqpldot$.
The equivalent plastic strain is chosen as the norm of the plastic strain tensor since it reflects deviatoric as well as volumetric deformations. 
\begin{align}
\eqpldot=\|\epltendot\|
\end{align}
If the associative flow and hardening laws are considered (see \cite{deSouzaNeto2008} p.~241) we have
\begin{align}
\Nten &= \pdiff{G}{\sten} = \pdiff{F}{\sten}\,, \quad \Aten=\pdiff{F}{\eqpl} \quad \text{and} \quad \Hten=1\,.
\end{align}
This implies that the partial derivatives
\begin{align}
\pdiff{\Hten}{\Aten} &= \nullvec \quad \text{and} \quad \pdiff{\Hten}{\sten} = \nullvec\,.
\end{align}
Under the assumption that the material parameters are constants Eq.~\eqref{eq:LinReturnMappingMatrix} simplifies to
\begin{align}
&
\begin{bmatrix}
\Dten + \Delta \plmult \pdiff{\Nten}{\sten} & \Nten \\[2ex]
\Nten & 0
\end{bmatrix}
\begin{bmatrix}
\text{d}\sten \\[2ex]
\text{d}\Delta\plmult
\end{bmatrix}
=
\begin{bmatrix}
\text{d}\eten^{\text{el trial}} \\[2ex]
0
\end{bmatrix}\,.
\label{eq:LinReturnMappingMatrix2}
\end{align}
For the general case of non-associate plastic flow and that hardening and plastic flow direction depending on the equivalent plastic strain $\eqpl$ and that $\eqpldot=\dot\plmult$ we have
\begin{align}
&
\begin{bmatrix}
\Dten + \Delta \plmult \pdiff{\Nten}{\sten} & \Nten \\[2ex]
\pdiff{F}{\sten} & \pdiff{F}{\eqpl}
\end{bmatrix}
\begin{bmatrix}
\text{d}\sten \\[2ex]
\text{d}\Delta\plmult
\end{bmatrix}
=
\begin{bmatrix}
\text{d}\eten^{\text{el trial}} \\[2ex]
0
\end{bmatrix}\,.
\label{eq:LinReturnMappingMatrix3}
\end{align}
To solve the equations \eqref{eq:LinReturnMappingMatrix2} or  \eqref{eq:LinReturnMappingMatrix3} a Newton scheme is applied. A vector of residuals is defined as
\begin{align}
\begin{bmatrix}
\Rvec^{\text{pl}}_{n+1} \\[1ex]
R^{F}_{n+1}
\end{bmatrix}
&=
\begin{bmatrix}
\eelten_{n+1} - \eelten_{n} - \Delta\eelten + \Delta \plmult \Nten_{n+1} \\[1ex]
F(\sten_{n+1},{\eqpl}_{n+1})
\end{bmatrix}
=
\begin{bmatrix}
\nullvec\\[1ex]
0
\end{bmatrix}\,,
\label{eq:ResVec}
\end{align}
and the Newton update for \eqref{eq:LinReturnMappingMatrix2} or  \eqref{eq:LinReturnMappingMatrix3} reads
\begin{align}
\begin{bmatrix}
\text{d}\sten_{n+1} \\[1ex]
\text{d}\Delta\plmult_{n+1}
\end{bmatrix}
&=
\AAten^{-1}_{n+1}
\begin{bmatrix}
-\Rvec^{\text{pl}}_{n+1} \\[1ex]
-R^{F}_{n+1}
\end{bmatrix}\,,
\label{eq:Newton1}
\end{align}
where $\AAten$ is the left most matrix defined in  \eqref{eq:LinReturnMappingMatrix2} or  \eqref{eq:LinReturnMappingMatrix3}.
The update of the variables $\sten$ and $\Delta\plmult$ is performed as
\begin{align}
\sten_{n+1} &= \sten_{n} + \text{d}\sten_{n+1}\,, \nonumber\\
\plmult_{n+1} & =\plmult_{n} + \text{d}\plmult_{n+1}\,. \label{eq:NewtonUpdate}
\end{align}
The Newton iterations are performed until the residuals become
\begin{align}
\|\Rvec^{\text{pl}}_{n+1}\| &< \text{tol} \quad \text{and} \quad \|R^F_{n+1}\| < \text{tol}\,.
\end{align}
The inverted matrix $A$ can be written using submatrices
\begin{align}
\AAten^{-1}
&=
\begin{bmatrix}
\Aten_{11} & \Aten_{12}\\
\Aten_{21} & \Aten_{22}
\end{bmatrix}\,.
\label{eq:Ainv}
\end{align}
Using \eqref{eq:Ainv}, linearization of \eqref{eq:Newton1} with respect to $\text{d}\eplten$ and considering that 
\begin{align}
\text{d}\eplten_{n+1} &= -\Cten^{-1} \, \text{d}\sten_{n+1}
= -\Dten \, \text{d} \sten_{n+1}\,,
\end{align}
one finds that
\begin{align}
\text{d}\sten_{n+1} &= \Cten_{\text{ep}} \text{d}\eten_{n+1} \quad \text{with} \quad \Cten_{\text{ep}}=\Aten_{11}\,,
\label{eq:ConsTangent}
\end{align}
where $\Cten_{\text{ep}}$ denotes the consistent tangent tensor.

\subsection{\label{sec:PI}Parameter Identification}
\subsubsection{\label{sec:PI:YS}Yield Surface}
The parameters for the modified model are obtained by analyzing a representative volume element (RVE) of a generic foam model as it is displayed in Fig.~\ref{fig:foam8}. A finite element model of the RVE is generated. The bulk material is assumed to be elastic-plastic described by a classical von Mises model with linear hardening. To determine the effective stresses and strains a homogenization approach as described by Malik et al.~\cite{Malik_AEM2022} is used. The FE analysis delivers for $n_l$ systematically varied load cases with $n_i$ load increments the data set
\begin{align}
\mathcal{D}^{\text{sim}}=\left\{\stenbar^{\text{sim}}_{(i,l)}\right\}\,,
\end{align}
where $\stenbar^{\text{sim}}_{(i,l)}$ denotes the homogenized stress tensor for load case $l$ and a corresponding homogenized equivalent plastic strain $\eqplbari$. The $i$ values for $\eqplbar$ are equally spaced in the range $[0 \dots \eqplbarmax]$ for every load case $l$. The $n_l$ load cases are defined in such a way that the stresses $\stenbar^{\text{sim}}_{(i,1\dots,n_l)}$ cover one of the six sections of the full yield surface for a corresponding equivalent plastic strain $\eqplbari$ as shown on the right side in Fig.~\ref{fig:foam8}. One section is representative, due to the symmetries of the yield surface. Further details regarding the sampling strategy can be found in \cite{Malik_AEM2022}. In other words, $\mathcal{D}_{i,:}$ represents a discretized yield surface for a given equivalent plastic strain. For each value of $\eqplbari$ a mean square error
\begin{align}
\mse\left(\pvec_i\right) &= \frac{1}{2\,n_l} \left\|\mathcal{D}^{\text{sim}}_{i,:} - \mathcal{D}^{\text{ana}}_{i,:}\left(\pvec_i\right)\right\|^2
\label{eq:mse}
\end{align}
is defined. The stresses in the set $\mathcal{D}^{\text{ana}}$ are obtained as
\begin{align}
\stenbar^{\text{ana}}_{(i,l)}\left(\pvec_i\right)=x\,\stenbar^{\text{sim}}_{(i,l)}\,,
\label{eq:xSsim}
\end{align}
and obey Eq.~\eqref{eq:ModEhlersYieldSurface} for a given parameter vector $\pvec_i=[\alpha_i,\beta_i,\gamma_i, \delta_i,\epsilon_i,\kappa_i,m_i]^T$, which has the length $n_p$. The scaling factor $x$ is determined numerically using a Newton iteration scheme
\begin{align}
x_{n+1} = x_n - \frac{F\left(x_n \stenbar^{\text{sim}},\pvec_i\right)}
{\pdiff{}{x}F\left(x_n \stenbar^{\text{sim}},\pvec_i\right)}
\quad \text{with} \quad x_0=1\,,
\label{eq:xNewton}
\end{align}
until $\left|F\left(x_n \stenbar^{\text{sim}},\pvec_i\right)\right| \le \text{tol}$ for each element in $\mathcal{D}^{\text{sim}}_{i,:}$. The index $i$ denotes the $i^{\text{th}}$ yield surface for $\eqplbari$ and $n$ the index for the Newton iterations. The elements in $\mathcal{D}^{\text{sim}}_{i,:}$ depend on the current parameters in $\pvec_i$. To find an optimal parameter set $\pvec_i^*$ for a yield surface with $\eqplbari$ the constrained minimization problem
\begin{align}
\pvec_i^* &= \argmin{\pvec_i} \left[\mse(\pvec_i)\right] \quad \text{subject to} \quad g_j(\pvec_i) \le 0\,; \quad j \, \in \, [1,2]
\label{eq:minimize}
\end{align}
has to be solved. The two inequality constraints $g_j$ are defined by the Eqs.~\eqref{eq:ConvexityCondFdev} and \eqref{eq:ConvexityCondFhyd} and are necessary to ensure the convexity of yield surface. A sequential least squares programming (SLSQP) algorithm is used to solve the constrained minimization problem defined in Eq.~\eqref{eq:minimize}. The algorithm requires the computation of the Jacobian of the objective function \eqref{eq:mse}. This can be done numerically using a finite difference scheme or much more efficiently analytically using
\begin{align}
\Jvec_i &= \frac{1}{n_l} \sum_{l=1}^{n_l} \left(\stenbar^{\text{sim}}_{(i,l)} - \stenbar^{\text{ana}}_{(i,l)} \right) : \pdiff{\stenbar^{\text{ana}}_{(i,l)}}{\pvec_i}\,,
\label{eq:Jacobian}
\end{align}
with
\begin{align}
\pdiff{\stenbar^{\text{ana}}_{(i,l)}}{\pvec_i}
&= 
\pdiff{\stenbar^{\text{ana}}_{(i,l)}}{x} * \pdiff{x}{\pvec_i} \,,
\label{eq:dSdP1}
\end{align}
considering the implicit derivative
\begin{align}
\pdiff{F}{\pvec_i} + \pdiff{F}{x} \pdiff{x}{\pvec_i} & \stackrel{!}{=} 0 \nonumber \\
\pdiff{F}{x} \pdiff{x}{\pvec_i} & = -\pdiff{F}{\pvec_i} \nonumber \\
\pdiff{x}{\pvec_i} & = -\left[ \pdiff{F}{x} \right]^{-1} \pdiff{F}{\pvec_i}\,,
\label{eq:dxdP}
\end{align}
such that
\begin{align}
\pdiff{\stenbar^{\text{ana}}_{(i,l)}}{\pvec_i}
& = - \pdiff{\stenbar^{\text{ana}}_{(i,l)}}{x} * \left[ \pdiff{F}{x} \right]^{-1} \pdiff{F}{\pvec_i}\,.
\label{eq:dSdP2}
\end{align}

%\pdiff{x}{F}
%\pdiff{F}{\pvec_i}
%=
%\pdiff{\stenbar^{\text{ana}}_{(i,l)}}{x} *
%\left[ \pdiff{F}{x} \right]^{-1}
%\pdiff{F}{\pvec_i}

The (*) product in Eq.~\eqref{eq:dSdP1} and \eqref{eq:dSdP2} denotes the outer product between a vector and a second order tensor, such that the result can be interpreted as a vector containing $n_p$ second-order tensors. The derivatives in \eqref{eq:dSdP2} are given in the appendix.
\subsubsection{\label{sec:PI:PFD}Plastic Flow Direction}
For the case of non-associated plastic flow, the direction $\Nten$ of the plastic strain is derived from a potential
\begin{align}
\Nten=\pdiff{G}{\sten} \quad \text{with} \quad G=\sqrt{J_2\left[1+\gamma_G A C\right]^{m_G} +\dfrac{1}{2} \alpha_G I_1^2}\,,
\label{eq:G}
\end{align}
which is a reduced version of the yield function, but with specific parameters $\alpha_G$, $\gamma_G$ and $m_G$. Again, these parameters may be scalar functions depending on internal variables (e.g.~$\eqpl$). The objective function to be minimized expresses the difference between the plastic flow directions determined by the analytical model and the numeric simulations using the FE model of the foam RVE.
\begin{align}
\mse_G(\pvec_{Gi})=\dfrac{1}{n_l}\sum_{n_l} \left( 1 - \dfrac{\Nten^{\text{ana}}_{(i,l)} : \Nten^{\text{sim}}_{(i,l)}}{\norm{\Nten^{\text{ana}}_{(i,l)}} \norm{\Nten^{\text{sim}}_{(i,l)}}} \right)
\label{eq:mseG}
\end{align}
The optimal parameters $\pvec_{Gi}^*$ are found by solving
\begin{align}
\pvec_{Gi}^* &= \argmin{\pvec_{Gi}} \left[\mse_G(\pvec_{Gi})\right] \,.
%\quad \text{subject to} \quad g_j(\pvec_i) \le 0\,; \quad j \, \in \, [1,2]
\label{eq:minimizeG}
\end{align}
The Jacobian of \eqref{eq:mseG} needed for the SLSQP minimization algorithm reads as
\begin{align}
\Jvec_{Gi}=\dfrac{-1}{n_l}\sum_{n_l} \pdiff{}{\pvec_{Gi}} \left( \dfrac{\Nten^{\text{ana}}_{(i,l)}}{\norm{\Nten^{\text{ana}}_{(i,l)}}} \right)\,. 
\label{eq:JacobianG}
\end{align}
The derivative of the normalized flow direction in the above equation is given in the appendix.
\section{Results, Applications, and Discussion}
\begin{figure}[htbp]
\centering
\includegraphics[width=0.4\columnwidth,trim={30 50 20 30},clip]{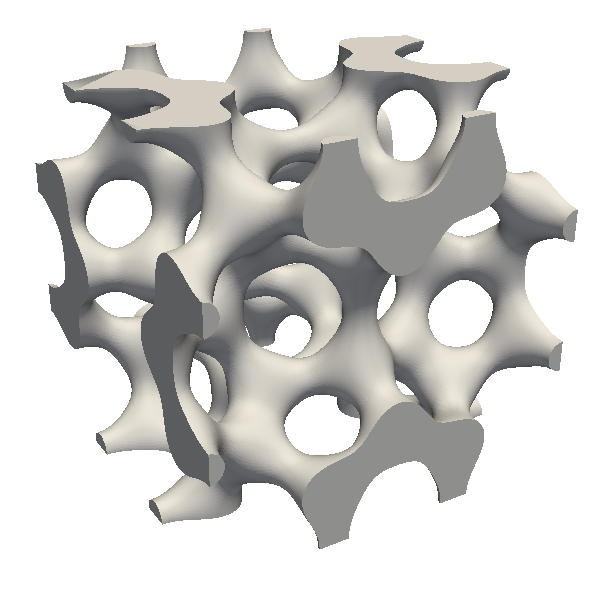}
\hfill
\includegraphics[width=0.55\columnwidth,trim={50 150 100 350},clip]{YS}
\caption{\label{fig:foam8}Wheire-Phelan foam structure RVE and its simulated yield surface for $\eqpl=0.2$.}
\end{figure}

The results given in this section are specific for the foam structure \cite{Abendroth_AEM2017,Malik_AEM2022} shown in Fig.~\ref{fig:foam8}. The local bulk material is assumed to be an isotropic elastic-plastic material with an elastic modulus $E_{\text{loc}}=10$\,GPa and a Poisson's ratio of $\nu_{\text{loc}}=0.3$. The local hardening law is linear with an initial yield stress of $\sigma_0=20$\,MPa and a linear hardening coefficient $\sigma_1=10$\,MPa. The relative density of the foam structure is $\bar{\rho}=20\%$ and the so-called strut shape factor is chosen as $k=1.0$ (see Abendroth et al.~\cite{Abendroth_AEM2017} for details).

\begin{figure}
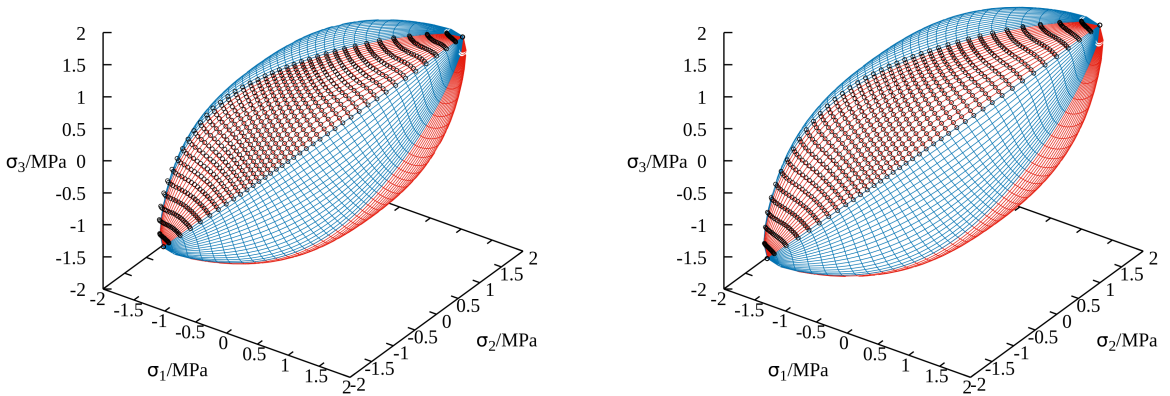

\centering
\includegraphics[width=0.49\columnwidth,trim={50 150 100 350},clip]{YSident0100}
\hfill
\includegraphics[width=0.49\columnwidth,trim={50 150 100 350},clip]{YSident0200}
\caption{\label{fig:YSident}Fitted yield surface for $\eqpl=0.1$ and $\eqpl=0.2$. The black dots represent the data set used for the parameter identification.}
\end{figure}

This structure is analyzed using DNSs with systematically varied stress states, which are defined as:
%Since it is assumed that the foam structure shows an isotropic elastic-plastic behavior, the stress states are defined in principal stress space.
\begin{align}
\stenbar &= \lambda \ntenbar \,, \\
\ntenbar &= \frac{I}{\sqrt{3}} \sin \alpha + \frac{\ntenhat}{\sqrt{2}} \cos \alpha \,, \\
\ntenhat &= \sum_{k=1}^3 e_k \Mten_k \quad \text{with} \quad \Mten_K=\Nten_k^{\stenbar} \otimes \Nten_k^{\stenbar} \,,
\end{align}
using
\begin{align}
\begin{bmatrix}
e_1\\e_2\\e_3
\end{bmatrix}
&=
\begin{bmatrix}
\cos \theta - \frac{\sin \theta}{\sqrt{3}}\\
\frac{2 \sin \theta}{\sqrt{3}}\\
-\cos \theta - \frac{\sin \theta}{\sqrt{3}}\\
\end{bmatrix}\,.
\end{align}
In the above equations $\lambda$ serves as a scaling factor, $\alpha \in [-\nicefrac{\pi}{2},\nicefrac{\pi}{2}]$ represents the angle between the deviatoric and hydrostatic stress direction. The eigenvalues $e_k$ of $\stenbar$ are defined using the Lode angle $\theta \in [-\nicefrac{\pi}{6},\nicefrac{\pi}{6}]$, and the three eigentensors $M_k$ are expressed by the eigendirections of the effective stress tensor $\Nten_k^{\stenbar}$. Since it is assumed that the foam structure shows an isotropic elastic-plastic behavior, the eigendirections are chosen to be equivalent to the unit directions of Euclidian space. The angles $\alpha$ and $\theta$ are varied uniformly in 39 and 19 steps within the given ranges. The scaling factor $\lambda$ is chosen such, that the equivalent effective strain $\eqplbarmax$ takes 50 equidistant levels between 0.5 and 25\%. In Fig.~\ref{fig:YSident} stress data are depicted exemplary for $\eqplbar=0.1$ and $\eqplbar=0.2$ as black dots superimposed to the corresponding yield surfaces.

\begin{figure}[htbp]
\includegraphics[width=0.49\columnwidth]{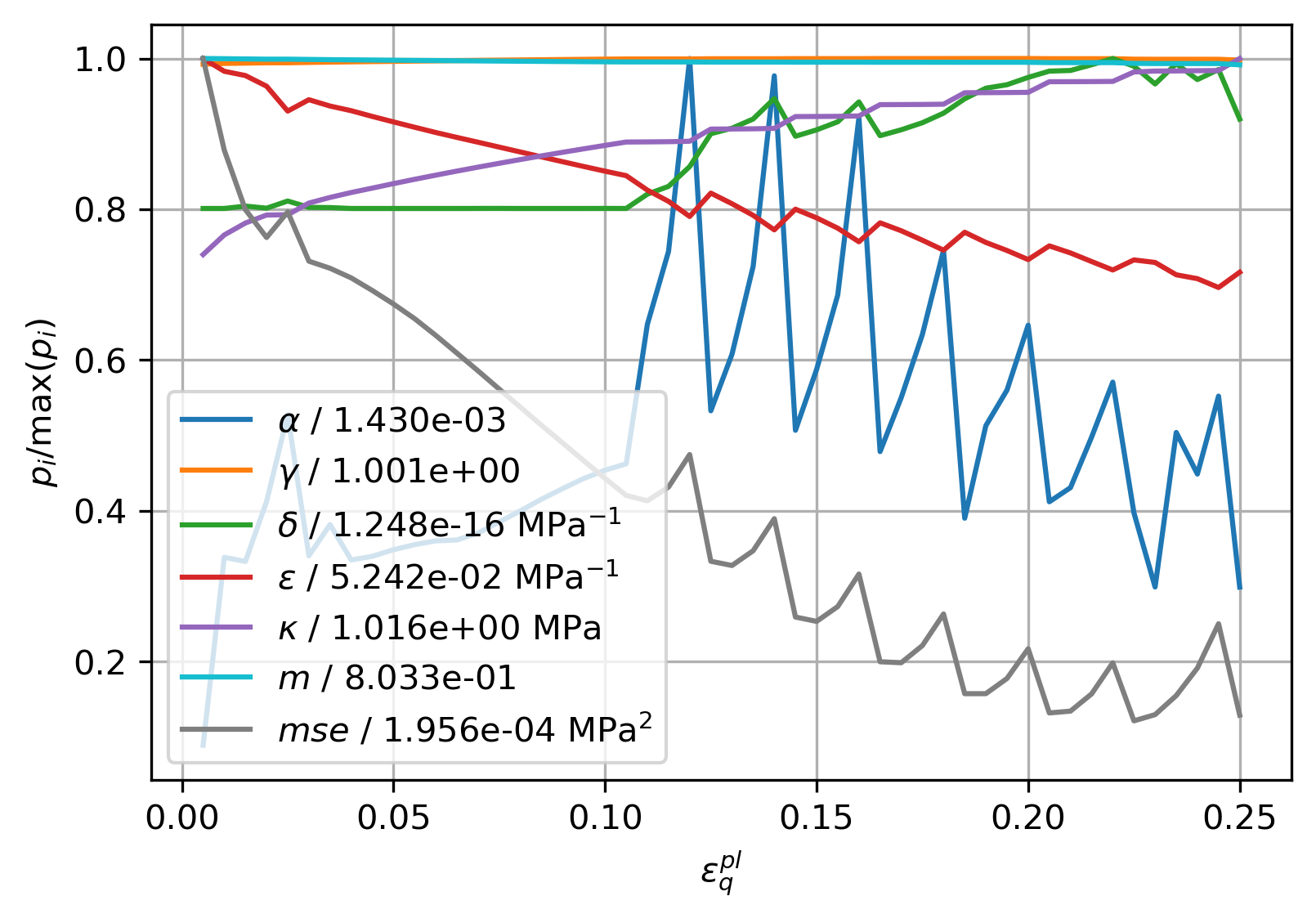}\hfill
\includegraphics[width=0.49\columnwidth]{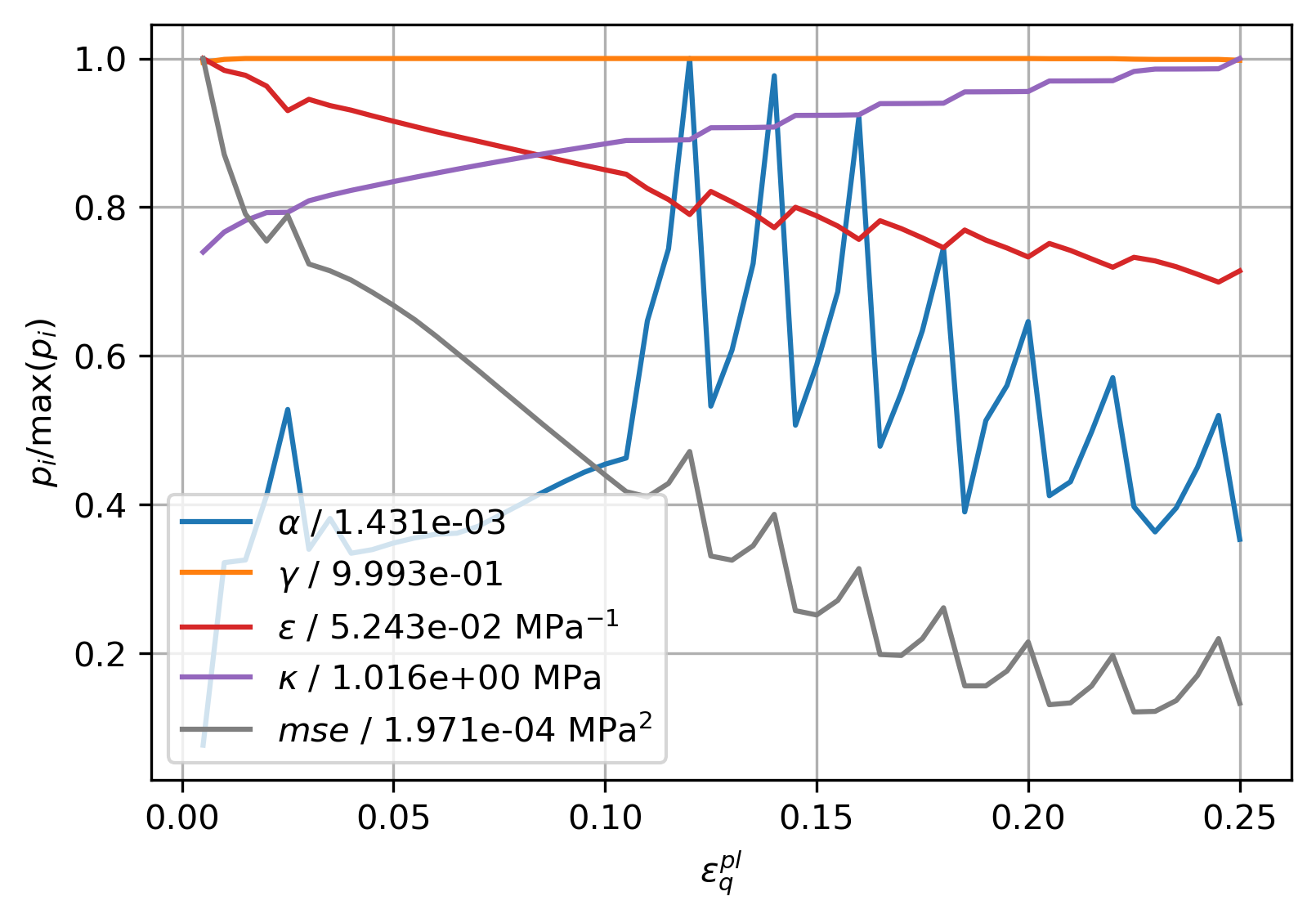}
\caption{\label{fig:ys-par-acdekm}Identified parameters depending on $\eqpl$: Left) The only fixed parameter is $\beta=0$. Right) With three fixed parameters $\beta=0$, $\delta=0$ and $m=0.8$.}
\end{figure}

Using that stress data for each level of equivalent effective strain a set of material parameters for the modified model is identified using the parameter identification procedure described in section \ref{sec:PI}. Since a small deformation setting and an isotropic local material model is considered, the effective yield surface is always point symmetric with respect to the origin, which implies that the parameter $\beta=0$. In the first attempt, all model parameters except $\beta$ have been identified. The values of these parameters are plotted over the effective equivalent plastic strain on the left-hand side in Fig.~\ref{fig:ys-par-acdekm}.

\begin{figure}[htbp]
\includegraphics[width=0.49\columnwidth]{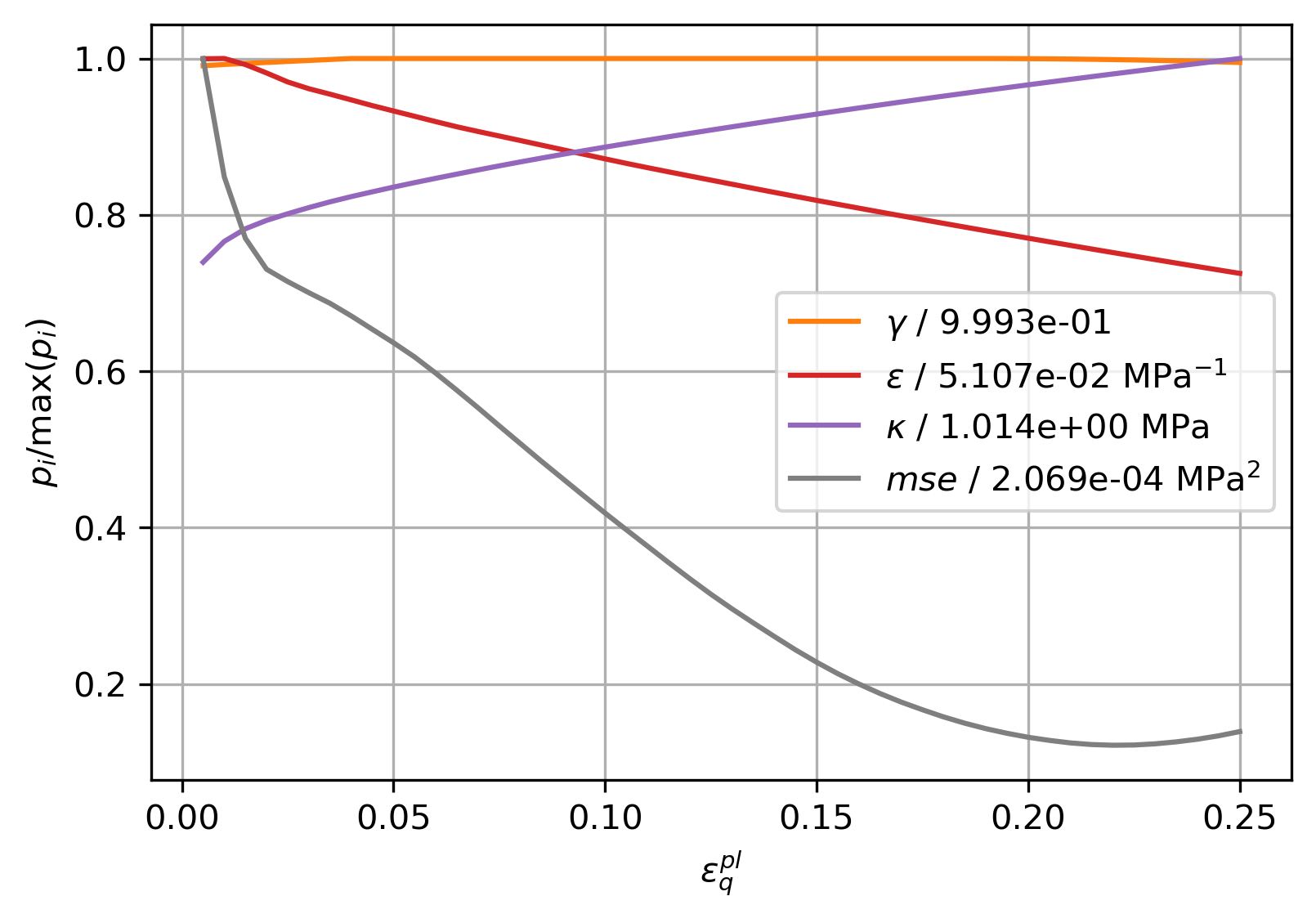}\hfill
\includegraphics[width=0.49\columnwidth]{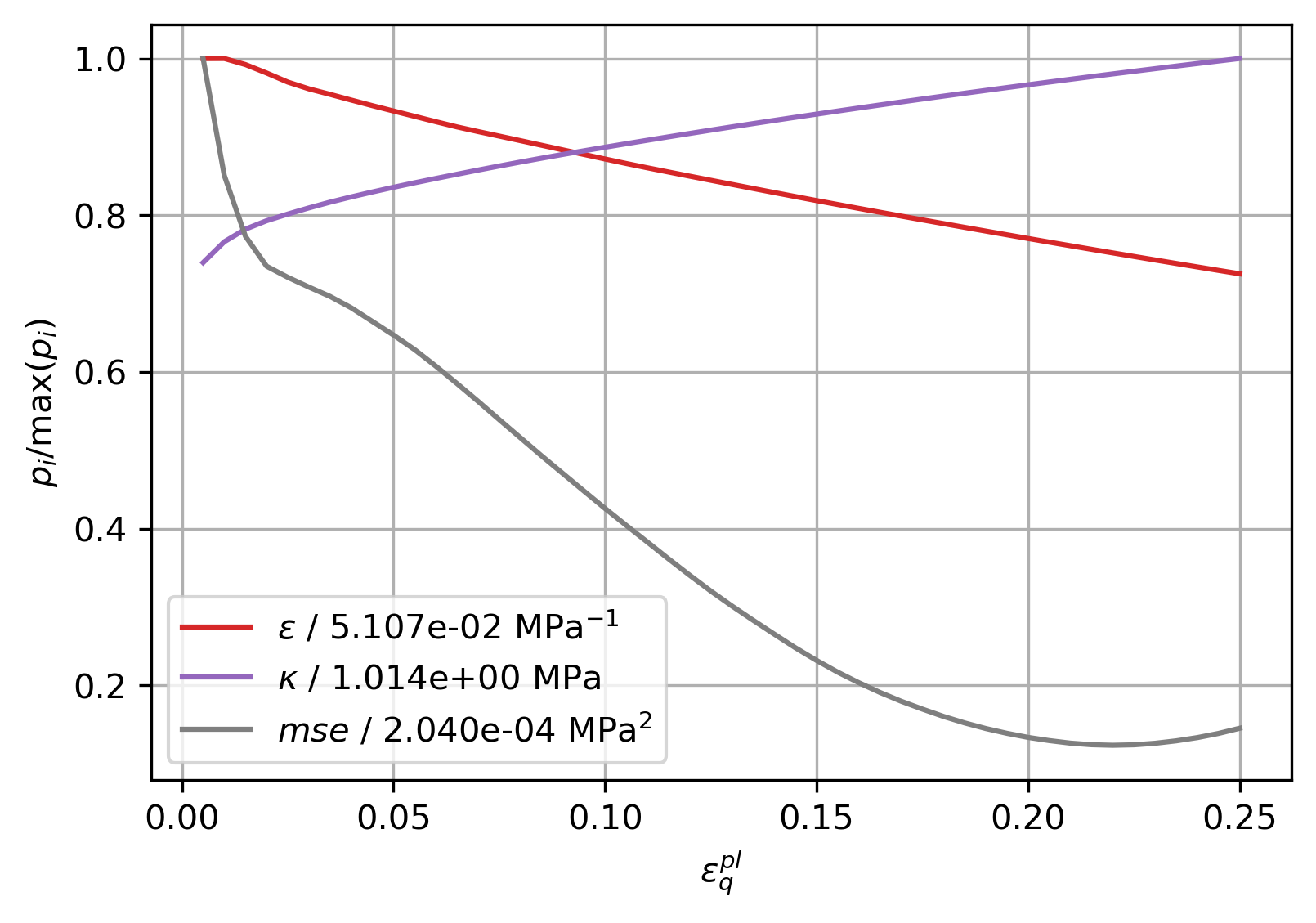}
\caption{\label{fig:ys-par-cek}Identified parameters depending on $\eqpl$: Left) The fixed parameters are $\alpha=7.068\cdot10^{-4}$, $\beta=0$, $\delta=0$, $m=0.8$. Right) The fixed parameters are $\alpha=7.068\cdot10^{-4}$, $\beta=0$, $\gamma=0.9981$, $\delta=0$, $m=0.8$.}
\end{figure}

Despite the fluctuations for $\alpha$ and $\epsilon$ it is observed that parameter $\delta$ is very close to zero and that $\gamma$ and $m$, which are the parameters defining the shape of the deviatoric cross-section, do not change significantly for increasing $\eqplbar$. The overall mean square error ($\mse$) has its maximum of $1.956\cdot10^{-4}$ MPa$^2$ at the lowest equivalent plastic strain value. That means, that the average error is about 0.014\,MPa, or less then 1\% of typical stress values, which can be interpreted as an excellent agreement between the analytical model and generic experiments. In a second parameter identification approach the parameters $\beta=0$, $\delta=0$, and $m=0.8$ remain fixed during optimization (see Fig.~\ref{fig:ys-par-acdekm} right). The remaining parameters show an almost identical behavior as in the first attempt and even the mean square error does not change significantly. A further reduction of the number of variable parameters leads to the results shown in Fig.~\ref{fig:ys-par-cek}.

On the left-hand side of Fig.~\ref{fig:ys-par-cek} the parameter values for $\gamma$, $\epsilon$, and $\kappa$ depending on $\eqplbar$ are plotted, and it is observed that also $\gamma$ does not change significantly. That, finally leads to the case where only the two parameters $\epsilon$ and $\kappa$ are used to describe the change of the yield surface shape during a plastic deformation process with very high accuracy as shown on the right panel in Fig.~\ref{fig:ys-par-cek}.
The dependency of the remaining two parameters $\epsilon$ and $\kappa$ on $\eqplbar$ can be described analytically using
\begin{align}
\epsilon &= \hat{\epsilon}(\eqplbar) = \epsilon_0 + \epsilon_1 \exp \left( -\epsilon_2 \eqplbar \right)\,, \label{eq:e_eqpl}\\
\kappa & = \hat{\kappa}(\eqplbar) = \kappa_0 + \kappa_1 \eqplbar + \kappa_2 \left[ 1 - \exp \left( -\kappa_3 \eqplbar \right) \right] + \kappa_4 \left[ 1 - \exp \left( -\kappa_5 \eqplbar \right) \right]\label{eq:k_eqpl}\,,
\end{align}
with
$\epsilon_0=0.02333$\,MPa$^{-1}$, $\epsilon_1=0.02818$\,MPa$^{-1}$, $\epsilon_2=2.841$, $\kappa_0=0.7022$\,MPa, $\kappa_1=0.6243$\,MPa, $\kappa_2=0.0752$\,MPa, $\kappa_3=150.1$, $\kappa_4=0.08512$\,MPa, $\kappa_5=11.96$. The corresponding fixed parameters are $\alpha=7.068\cdot10^{-4}$, $\beta=0$, $\gamma=0.9981$, $\delta=0$, and  $m=0.8$

\begin{figure}[htbp]
\centering
\includegraphics[width=0.49\columnwidth]{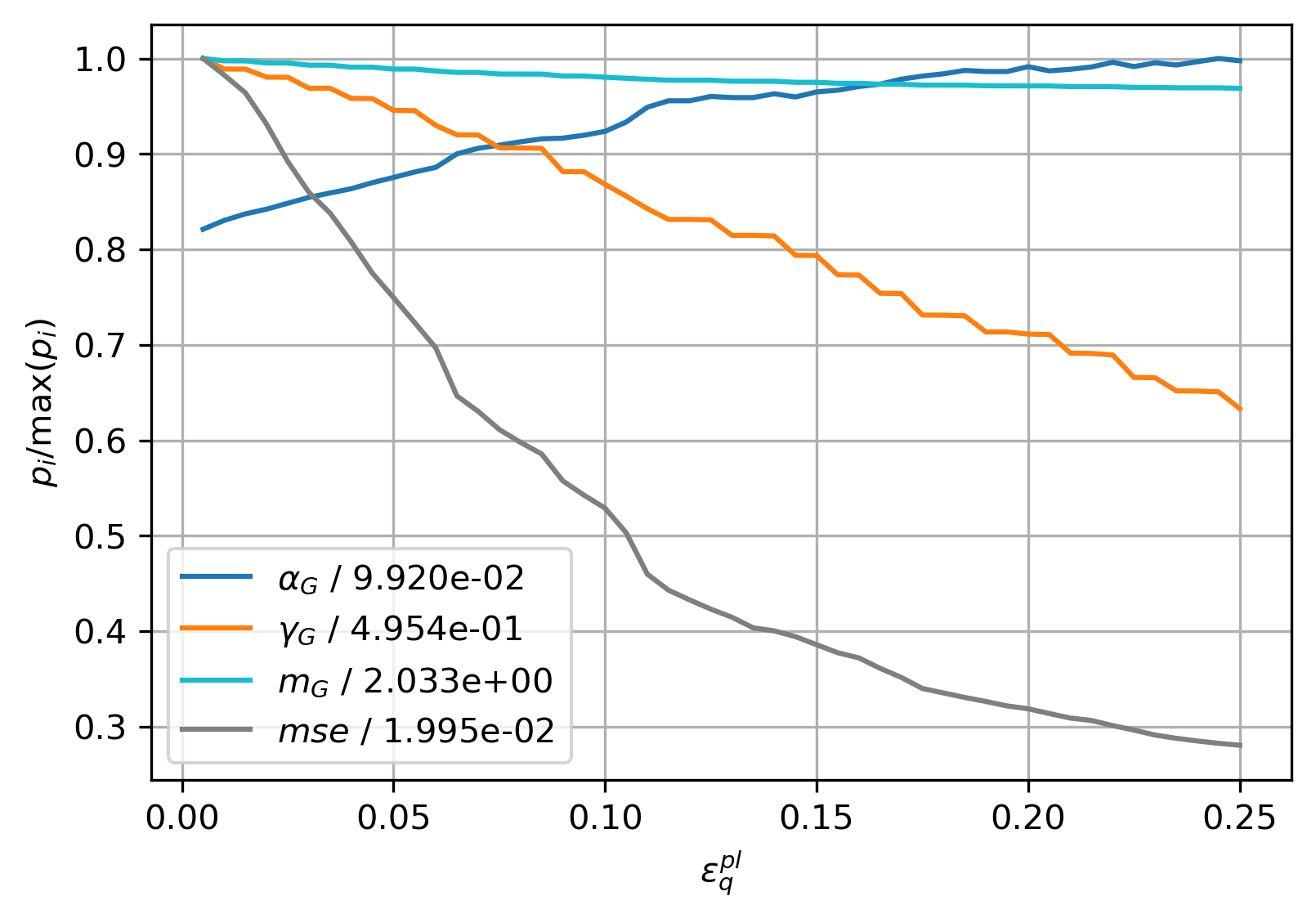}
\caption{\label{fig:agcgmg}Identified parameters for the flow potential depending on $\eqpl$.}
\end{figure}

The dependencies of three dimensionless parameters of the flow potential $G$ on the equivalent plastic strain are depicted in Fig.~\ref{fig:agcgmg}. Also here, the parameters can be expressed by scalar functions.
\begin{align}
\alpha_G & = \hat{\alpha_G}(\eqplbar) = \alpha_{G0} + \alpha_{G1} \eqplbar + \alpha_{G2} \left[ 1 - \exp \left( -\alpha_{G3} \eqplbar \right) \right]\label{eq:a_eqpl}\\
\gamma_G & = \hat{\gamma_G}(\eqplbar) = \gamma_{G0} + \gamma_{G1} \eqplbar + \gamma_{G2} \left[ 1 - \exp \left( -\gamma_{G3} \eqplbar \right) \right]\label{eq:c_eqpl}\\
m_G & = \hat{m_G}(\eqplbar) = m_{G0} + m_{G1} \eqplbar + m_{G2} \left[ 1 - \exp \left( -m_{G3} \eqplbar \right) \right]\label{eq:m_eqpl}
\end{align}
The corresponding parameters are $\alpha_{G0}=0.08002$, $\alpha_{G1}=0.0$, $\alpha_{G2}=0.02243$, $\alpha_{G3}=8.065$, $\gamma_{G0}=0.5049$, $\gamma_{G1}=-0.7625$, $\gamma_{G2}=-0.2236$, $\gamma_{G3}=0.01241$, $m_{G0}=2.036$, $m_{G1}=0.1791$, $m_{G2}=-0.1561$, and $m_{G3}=4.886$. The mean square error reduces with increasing $\eqpl$ from $0.01995$ down to $0.005594$. 

\begin{figure}[htbp]
\includegraphics[width=0.49\columnwidth]{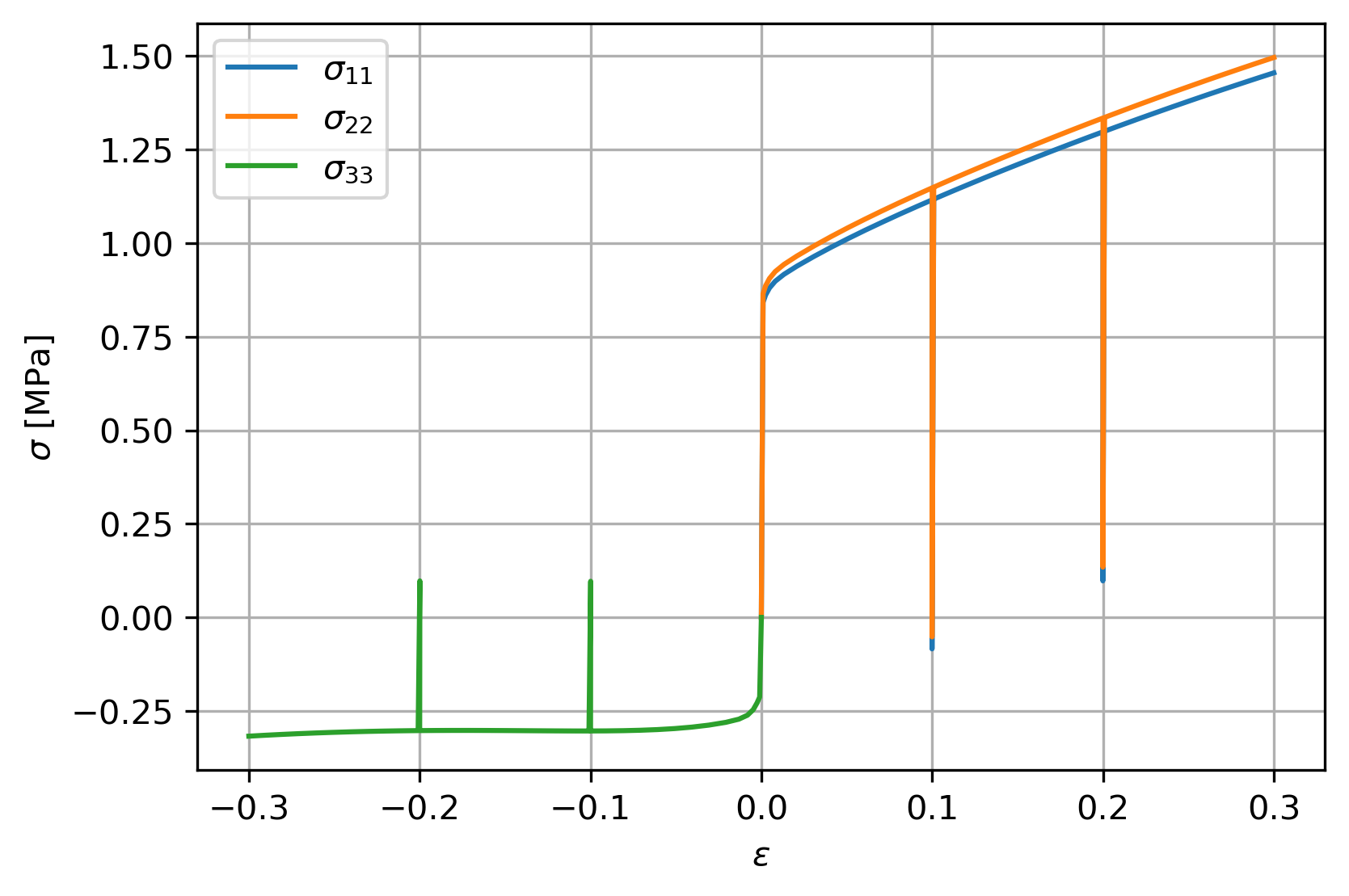}\hfill
\includegraphics[width=0.49\columnwidth]{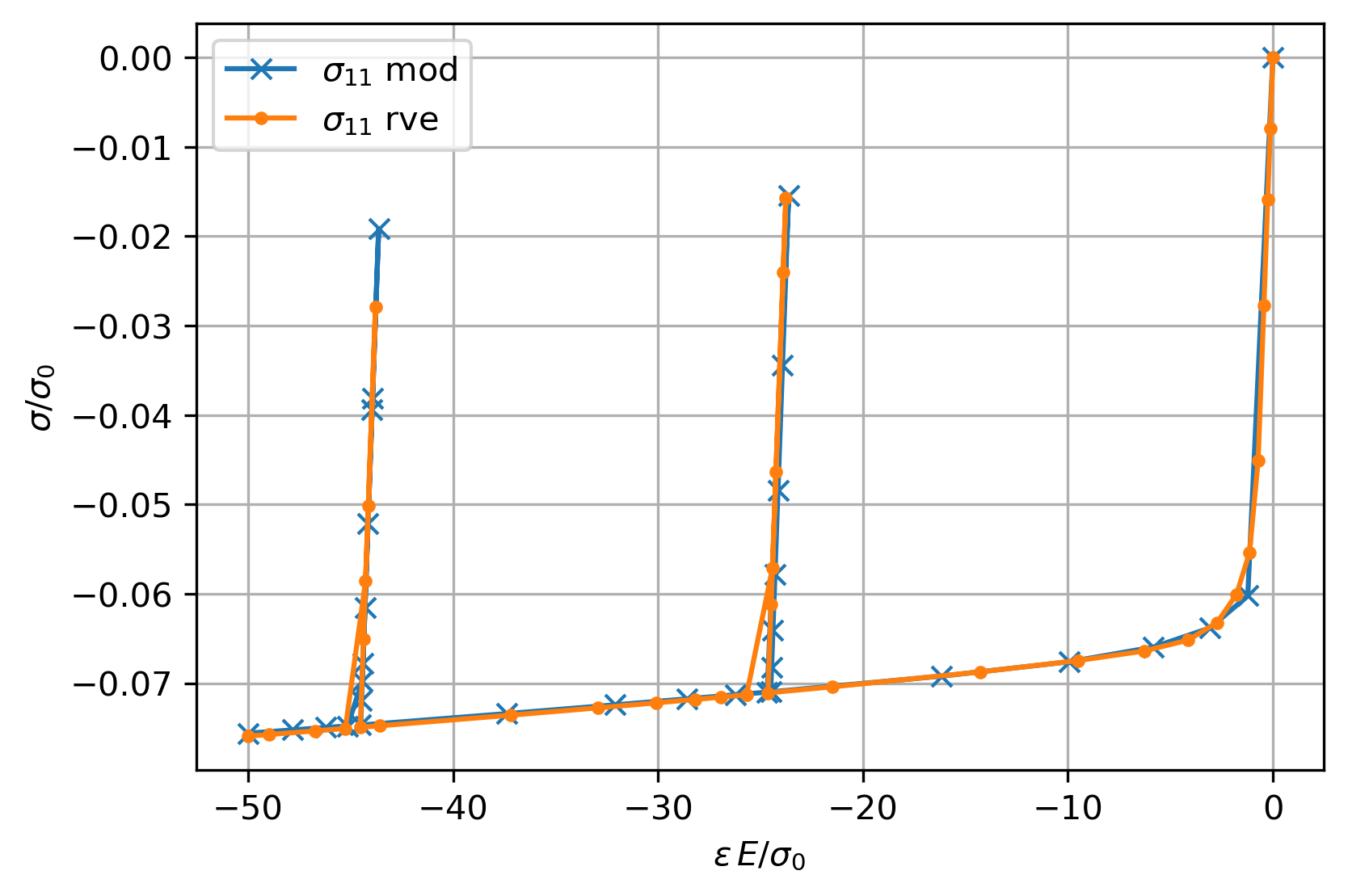}
\caption{\label{fig:strain-stress-1-2}Stress-strain plots for left) an uniaxial positive strain ($\bar{\varepsilon}_{11}=0.1$, $\bar{\varepsilon}_{22}=0.0$, $\bar{\varepsilon}_{33}=0.0$) and right) an uniaxial negative strain ($\bar{\varepsilon}_{11}=-0.1$, $\bar{\varepsilon}_{22}=0.0$, $\bar{\varepsilon}_{33}=0.0$) controlled load case with two unloading slopes each.}
\end{figure}

\begin{figure}[htbp]
\includegraphics[width=0.49\columnwidth]{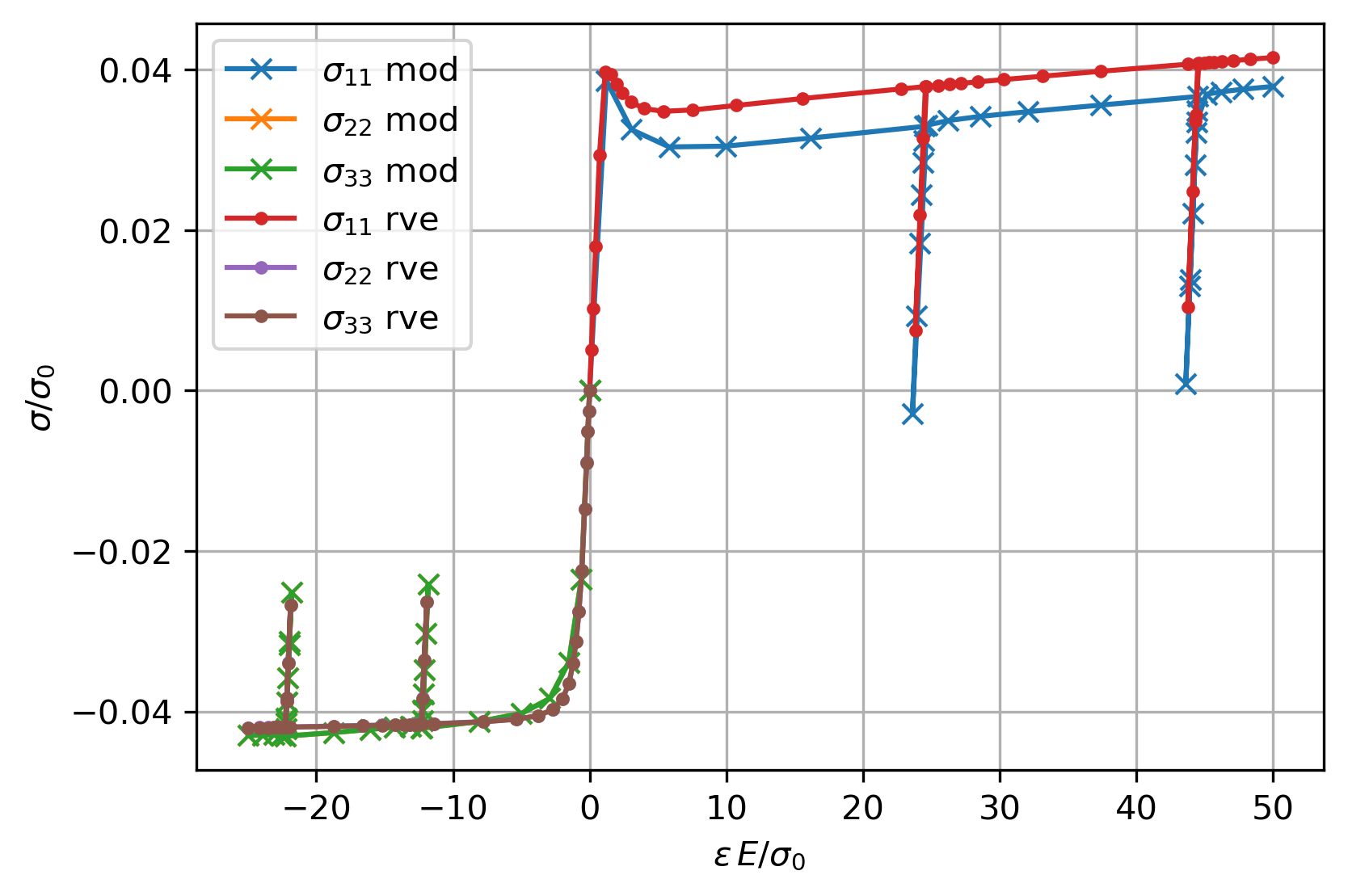}\hfill
\includegraphics[width=0.49\columnwidth]{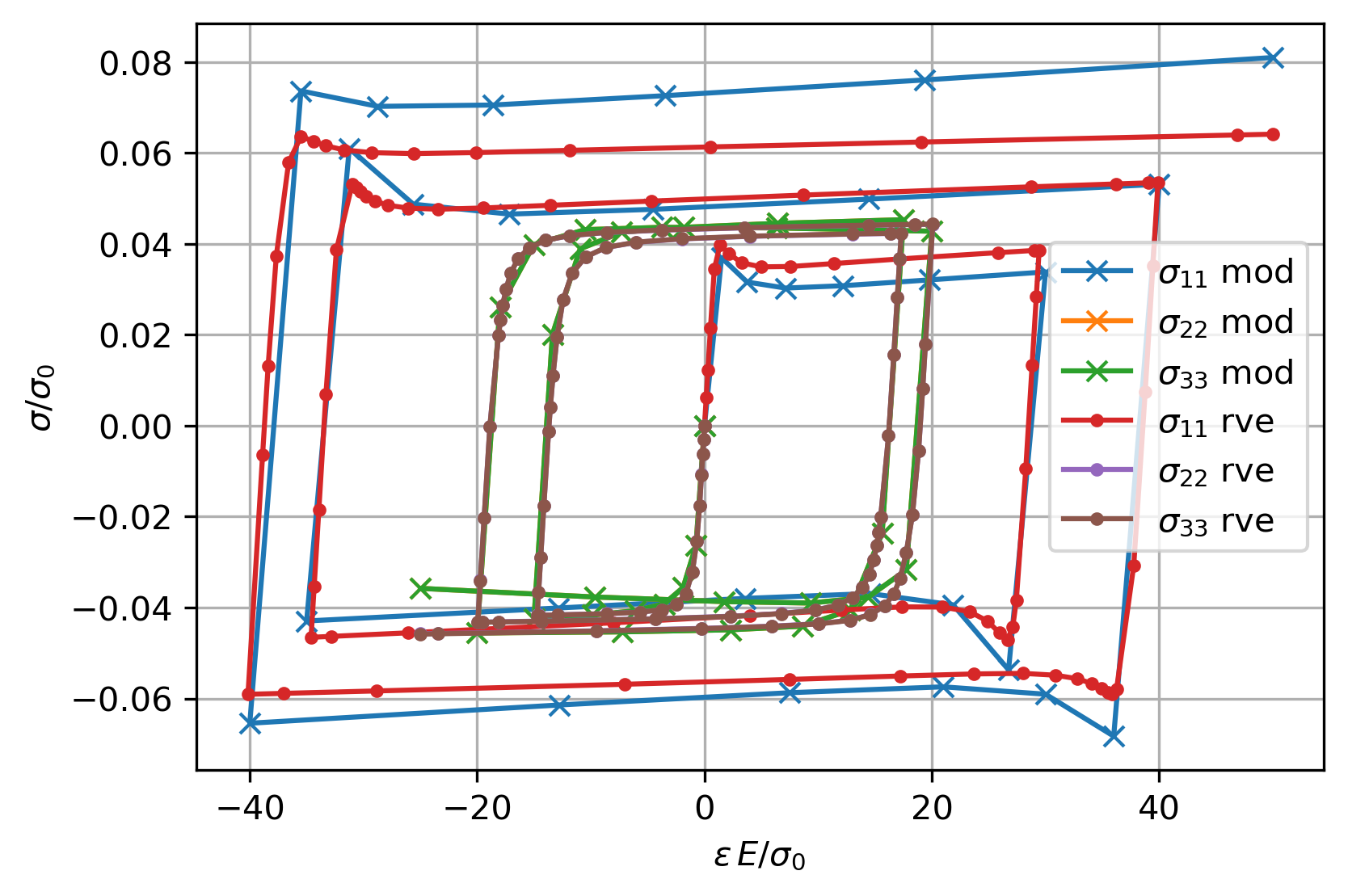}
\caption{\label{fig:strain-stress-dev-cycl}left) Stress-strain plots for a strain controlled deviatoric ($\bar{\varepsilon}_{11}=-2\bar{\varepsilon}_{22}=-2\bar{\varepsilon}_{33}$) load case. right) Stress-strain plots for a strain controlled ($\bar{\varepsilon}_{11}=-2\bar{\varepsilon}_{22}=-2\bar{\varepsilon}_{33}$) cyclic load case.}
\end{figure}

The above-identified parameters have been used to predict the stress-strain response for various load cases. In Fig.~\ref{fig:strain-stress-1-2} the predictions of the model for uniaxial positive and negative load cases are compared with the results from direct numerical simulations of the Wheire-Phelan RVE shown in Fig.~\ref{fig:foam8}. Here, an excellent agreement between model prediction and data from RVE simulations is observed. 
In Fig.~\ref{fig:strain-stress-dev-cycl} left) A deviatoric loading case is investigated. For the given strain state it is observed that $\sigma_{11}$ reaches a peak at the end of the elastic region followed by a short stress drop before the usual strain hardening starts, which is also reflected by the material model. The further evolution of stresses is slightly more deviating than for the uniaxial load cases but still accurate enough for most engineering applications.
As a final comparison, a cyclic load with an increasing strain amplitude in each half cycle is investigated and visualized in Fig.~\ref{fig:strain-stress-dev-cycl} right). Here, the stress-strain curve of the developed material model deviates increasingly from the data with each cycle. Since the model considers only isotropic hardening effects, possible kinematic hardening of the RVE is not reflected. Nevertheless, the model is able to approximate the general material behavior with reasonable accuracy.

\section{Summary, Discussion, and Outlook}
A new constitutive model for the elastic-plastic behavior of foams or other porous structures has been presented. The model is a modification of Ehlers' model \cite{Ehlers1995} and can especially describe the change of the orientation of the triangular-shaped deviatoric yield surface cross-section, depending on the hydrostatic stress state. The model is formulated using a consistent thermodynamic framework by de Souza Neto \cite{deSouzaNeto2008}. A general return algorithm is used to solve the constitutive equations. A comprehensive appendix contains all necessary partial derivatives needed for the implementation of the model into finite element codes and the application of the parameter identification procedure.

The model parameters are identified using a constraint parameter identification procedure, where data from direct numerical simulations of a representative volume element of a generic foam model are used as a substitution for experimental data. The parameter identification procedure is not generally restricted to generic data, sufficient experimental data can also be used. Care must be taken for the choice of the parameters since their exist certain constraints to ensure the overall convexity of the yield surface.

The predictions of the model are compared with data generated by DNS of the generic foam model and show an excellent agreement. The authors believe that this model is capable to describe a wide range of foam structures. The structure of the thermodynamical framework allows certain extensions such as the application for finite deformations and the use of additional internal variables. Also, other failure criteria like the Bigoni-Piccolroaz criteria \cite{Bigoni2004} or the GMM criteria described in Bolchoun et al.~\cite{BolchounKolupaevAltenbach2011} could be implemented as yield surfaces or flow potentials using the same framework.

The current version of the model is implemented in a small deformation setting. It does not allow the simulation of foam compaction processes, where different elements of the mesostructure come into contact. Also, any anisotropic properties that can be observed in some real foams are not considered here. But, these deficiencies could be eliminated in future versions, on which the authors are currently working on.

\bibliographystyle{plain}
\bibliography{PlasEhlersModLiterature}

\begin{thebibliography}{10}

\bibitem{Abendroth_AEM2017}
M.~Abendroth, E.~Werzner, C.~Settgast, and S.~Ray.
\newblock An approach toward numerical investigation of the mechanical behavior
  of ceramic foams during metal melt filtration processes.
\newblock {\em Advanced Engineering Materials}, 19(9):1700080, 2017.

\bibitem{Altenbach2014}
H.~Altenbach, A.~Bolchoun, and V.~A. Kolupaev.
\newblock {\em Phenomenological Yield and Failure Criteria}, pages 49--152.
\newblock Springer Berlin Heidelberg, Berlin, Heidelberg, 2014.

\bibitem{Ashby2006}
M.~F. Ashby.
\newblock The properties of foams and lattices.
\newblock {\em Philosophical Transactions of the Royal Society A: Mathematical,
  Physical and Engineering Sciences}, 364(1838):15--30, 2006.

\bibitem{Barlat1991}
F.~Barlat, D.~J. Lege, and J.~C. Brem.
\newblock A six-component yield function for anisotropic materials.
\newblock {\em International Journal of Plasticity}, 7(7):693--712, 1991.

\bibitem{Bigoni2004}
D.~Bigoni and A.~Piccolroaz.
\newblock Yield criteria for quasibrittle and frictional materials.
\newblock {\em International Journal of Solids and Structures},
  41(11):2855--2878, 2004.

\bibitem{Bilkhu1993}
S.~S. Bilkhu, M.~Founas, and G.~S. Nusholtz.
\newblock Material modeling of structural foams in finite element analysis
  using compressive uniaxial and triaxial data.
\newblock In {\em International Congress \& Exposition}. SAE International, mar
  1993.

\bibitem{BolchounKolupaevAltenbach2011}
A.~Bolchoun, V.~A. Kolupaev, and H.~Altenbach.
\newblock Konvexe und nichtkonvexe fließflächen.
\newblock {\em Forschung im Ingenieurwesen}, 75:73--92, 2011.

\bibitem{deSouzaNeto2008}
E.~A. de~Souza~Neto, D.~Peric, and D.~R.~J. Owen.
\newblock {\em Computational Methods for Plasticity}.
\newblock John Wiley \& Sons, LtD, 2008.

\bibitem{Demiray2007}
S.~Demiray, W.~Becker, and J.~Hohe.
\newblock Numerical determination of initial and subsequent yield surfaces of
  open-celled model foams.
\newblock {\em International Journal of Solids and Structures},
  44(7):2093--2108, 2007.

\bibitem{DeshpandeFleck2000}
V.~S. Deshpande and N.~A. Fleck.
\newblock Isotropic constitutive models for metallic foams.
\newblock {\em Journal of the Mechanics and Physics of Solids},
  48(6):1253--1283, 2000.

\bibitem{Ehlers1995}
W.~Ehlers.
\newblock A single-surface yield function for geomaterials.
\newblock {\em Archive of Applied Mechanics}, 65(4):246--259, 1995.

\bibitem{EhlersAvci2012}
W.~Ehlers and O.~Avci.
\newblock Stress-dependent hardening and failure surfaces of dry sand.
\newblock {\em International Journal for Numerical and Analytical Methods in
  Geomechanics}, 37(8):787--809, 2012.

\bibitem{FahlbuschGrenestedtBecker2016}
N.-C. Fahlbusch, J.L. Grenestedt, and W.~Becker.
\newblock Effective failure behavior of an analytical and a numerical model for
  closed-cell foams.
\newblock {\em International Journal of Solids and Structures}, 97-98:417--430,
  2016.

\bibitem{FlorenceSab2005}
C.~Florence and K.~Sab.
\newblock Overall ultimate yield surface of periodic tetrakaidecahedral lattice
  with non-symmetric material distribution.
\newblock {\em Journal of Materials Science}, 40(22):5883--5892, 2005.

\bibitem{GibsonAshby1989}
L.~J. Gibson, M.~F. Ashby, J.~Zhang, and T.~C. Triantafillou.
\newblock Failure surfaces for cellular materials under multiaxial
  loads—i.modelling.
\newblock {\em International Journal of Mechanical Sciences}, 31(9):635--663,
  1989.

\bibitem{JungDiebels2018}
A.~Jung and S.~Diebels.
\newblock Yield surfaces for solid foams: A review on experimental
  characterization and modeling.
\newblock {\em GAMM-Mitteilungen}, 41(2):e201800002, 2018.

\bibitem{Laroussi2002}
M.~Laroussi, K.~Sab, and A.~Alaoui.
\newblock Foam mechanics: nonlinear response of an elastic 3d-periodic
  microstructure.
\newblock {\em International Journal of Solids and Structures},
  39(13):3599--3623, 2002.

\bibitem{Luxner2007}
M.~H. Luxner, J.~Stampfl, and H.~E. Pettermann.
\newblock Numerical simulations of 3d open cell structures – influence of
  structural irregularities on elasto-plasticity and deformation localization.
\newblock {\em International Journal of Solids and Structures},
  44(9):2990--3003, 2007.

\bibitem{Malik_AEM2022}
A.~Malik, M.~Abendroth, G.~Huetter, and B.~Kiefer.
\newblock A hybrid approach employing neural networks to simulate the
  elasto-plastic deformation behavior of 3d-foam structures.
\newblock {\em Advanced Engineering Materials}, 24(2):2100641, 2022.

\bibitem{Nusholtz1996}
G.~S. Nusholtz, S.~Bilkhu, M.~Founas, K.~Uduma, and P.~A. DeBois.
\newblock Impact response of foam: The effect of the state of stress.
\newblock In {\em 40th Stapp Car Crash Conference (1996)}. SAE International,
  nov 1996.

\bibitem{Oechsner2010}
A~{\"O}chsner.
\newblock {\em Plasticity of Three-Dimensional Foams}, pages 107--166.
\newblock Springer Vienna, Vienna, 2010.

\bibitem{Settgast_MOM2019}
C.~Settgast, M.~Abendroth, and M.~Kuna.
\newblock Constitutive modeling of plastic deformation behavior of open-cell
  foam structures using neural networks.
\newblock {\em Mechanics of Materials}, 131:1--10, 2019.

\bibitem{Settgast_IJP2020}
C.~Settgast, G.~H{\"u}tter, M.~Kuna, and M.~Abendroth.
\newblock A hybrid approach to simulate the homogenized irreversible
  elastic--plastic deformations and damage of foams by neural networks.
\newblock {\em International Journal of Plasticity}, 126:102624, 2020.

\bibitem{Storm2016}
J.~Storm, M.~Abendroth, and M.~Kuna.
\newblock Numerical and analytical solutions for anisotropic yield surfaces of
  the open-cell kelvin foam.
\newblock {\em International Journal of Mechanical Sciences}, 105:70--82, 2016.

\bibitem{StormAEM2015}
J.~Storm, M.~Abendroth, D.~Zhang, and M.~Kuna.
\newblock Geometry dependent effective elastic properties of open-cell foams
  based on kelvin cell models.
\newblock {\em Advanced Engineering Materials}, 15(12):1292--1298, 2013.

\bibitem{TsaiWu1971}
S.~W. Tsai and E.~M. Wu.
\newblock A general theory of strength for anisotropic materials.
\newblock {\em Journal of Composite Materials}, 5(1):58--80, 1971.

\bibitem{Wang2005}
A.-J. Wang and D.~L. McDowell.
\newblock Yield surfaces of various periodic metal honeycombs at intermediate
  relative density.
\newblock {\em International Journal of Plasticity}, 21(2):285--320, 2005.

\bibitem{Wang2006}
D.-A. Wang and J.~Pan.
\newblock A non-quadratic yield function for polymeric foams.
\newblock {\em International Journal of Plasticity}, 22(3):434--458, 2006.

\bibitem{Zhang2015}
D.~Zhang, M.~Abendroth, M.~Kuna, and J.~Storm.
\newblock Multi-axial brittle failure criterion using weibull stress for open
  kelvin cell foams.
\newblock {\em International Journal of Solids and Structures}, 75-76:1--11,
  2015.

\end{thebibliography}

\appendix
\section{Appendix}
This appendix presents derivatives of the yield function and/or the flow potential, which are necessary for the implementation of the model into a finite element code. Furthermore, the derivatives necessary for the parameter identification procedure are given.
\subsection{Necessary derivatives for the implementation of the model}
\subsubsection{Derivatives of the yield function}
We start with subsequent substitutions of terms in Eq.~\eqref{eq:ModEhlersYieldSurface}.
\begin{align}
A &= \frac{\tr{\Nten}}{\sqrt{3}}  \label{eq:A}\\
B &= \frac{J_3}{J_2^{\nicefrac{3}{2}}}  \label{eq:B}\\
C &= A \cdot B  \label{eq:C}\\
D &= \left(1 + \gamma C\right)^m  \label{eq:D}\\
E &= J_2 \, D  \label{eq:E}\\
W &= \sqrt{E+\frac{1}{2}\alpha I_1^2 + \delta^2 I_1^4} \label{eq:W}
\end{align}
Therewith, Eq.~\eqref{eq:ModEhlersYieldSurface} can be expressed as
\begin{align}
F &= W + \beta I_1 + \epsilon I_1^2 - k\,.
\label{eq:F}
\end{align}
The first derivatives of the stress invariants with respect to the symmetric Cauchy stress tensor $\sten$ reads
\begin{align}
\pdiff{I_1}{\sten}=\Iten\,, \quad \pdiff{J_2}{\sten}=\sdev\,, \quad \text{and} \quad \pdiff{J_3}{\sten}=\sdev\sdev - \frac{2}{3}J_2\Iten\,.
\label{eq:InvDer}
\end{align}
In Eqs.~\eqref{eq:InvDer}\textsubscript{2,3} the symbol $\sdev$ denotes the deviator of the symmetric Cauchy stress tensor.
\begin{align}
\sdev &= \sten - \frac{1}{3} \tr{\sten} \Iten = \IItendev : \sten
\label{eq:StressDeviator} 
\end{align} 
The derivatives of the powers of the first invariants in \eqref{eq:W} and \eqref{eq:F} are given by
\begin{align}
\pdiff{I_1^2}{\sten}=2I_1\Iten \quad \text{and} \quad \pdiff{I_1^4}{\sten}=4I_1^3\Iten\,.
\label{InvPowDer}
\end{align}
Using \eqref{eq:InvDer} and \eqref{InvPowDer} the derivatives of \eqref{eq:A} -- \eqref{eq:W} with respect to $\sten$ become
\begin{align}
\pdiff{A}{\sten} &= \frac{\left( \Iten \norm{\sten}^2 - I_1 \sten\right)}{\sqrt{3}\norm{\sten}^3}
 \label{eq:dAdS}\\[1ex]
\pdiff{B}{\sten} &= \frac{2 J_2 \pdiff{J_3}{\sten} - 3 J_3 \pdiff{J_2}{\sten}}{2 J_2^{\nicefrac{5}{3}}} \label{eq:dBdS}\\[1ex]
\pdiff{C}{\sten} &= A \pdiff{B}{\sten} + B \pdiff{A}{\sten} \label{eq:dCdS}\\[1ex]
\pdiff{D}{\sten} &= \gamma\,m \pdiff{C}{\sten}\left(1 + \gamma\,C\right)^{m-1} \label{eq:dDdS}\\[1ex]
\pdiff{E}{\sten} &= J_2 \pdiff{D}{\sten} + D \pdiff{J_2}{\sten} \label{eq:dEdS}\\[1ex]
\pdiff{W}{\sten} &= \frac{\pdiff{E}{\sten} + \alpha I_1 \Iten + 4 \delta^2 I_1^3 \Iten}{2 W} \label{eq:dWdS}\,.
\end{align}
Having these derivatives one finally gets
\begin{align}
\pdiff{F}{\sten} &= \pdiff{W}{\sten} + \beta \Iten + 2 \epsilon I_1 \Iten\,.
\end{align}
If a Newton algorithm is considered for solving the constitutive equations the second derivatives of $F$ and $G$ with respect to $\sten$ are required. Here, it is important to keep in mind that $\sten$ is symmetric. The resulting following derivatives are symmetric fourth-order tensors. Several forth order unit tensors are needed for the formulation of the second derivatives. The general forth order unit tensor and its transpose can be derived from an unsymmetric tensor $\Aten$.
\begin{align}
\IIten &= \pdiff{\Aten}{\Aten}, \quad \IItenT=\pdiff{\Aten^T}{\Aten}
\label{eq:II}
\end{align}
The symmetric fourth-order unit tensor is therewith defined as
\begin{align}
\IItensym &= \pdiff{\Aten^{\text{sym}}}{\Aten} 
= \pdiff{\frac{1}{2}\left(\Aten+\Aten^T\right)}{\Aten}
= \frac{1}{2	}\left(\IIten+\IItenT\right)
\label{eq:IIsym}
\end{align}
and reads in index notation as
\begin{align}
\IItensym_{ijkl} &= \frac{1}{2}\left(\delta_{ik}\delta_{jl} + \delta_{il}\delta_{jk}\right)\,.
\label{eq:IIsymidx}
\end{align}
The volumetric fourth-order tensor is defined as
\begin{align}
\IItenvol &= \pdiff{\Aten^{\text{vol}}}{\Aten}=\frac{1}{3}\Iten \otimes \Iten
\label{eq:IIvol}
\end{align}
and reads in index notation
\begin{align}
\IItenvol_{ijkl} &= \frac{1}{3}\left(\delta_{ij}\delta_{kl}\right)\,.
\label{eq:IIvolidx}
\end{align}
The deviatoric forth order unit tensor is finally defined as
\begin{align}
\IItendev=\pdiff{\Aten^{\text{dev}}}{\Aten}=\IItensym-\IItenvol\,.
\label{eq:IIdev}
\end{align}
With the fourth-order unit tensors at hand, the second derivatives for the stress invariants can be formulated as follows.
\begin{align}
\pddiff{I_1}{\sten} &= \Nullten \label{eq:d2I1dS2}\\[1ex]
\pddiff{J_2}{\sten} &= \pdiff{\sdev}{\sten}=\IItendev \label{eq:d2J2dS2}\\[1ex]
\pddiff{J_3}{\sten} &= \pdiff{\left( \sdev\sdev - \frac{2}{3}J_2\Iten \right)}{\sten}= \IItendev \sdev  +  \sdev \, \IItendev - \frac{2}{3} \Iten \otimes \sdev \label{eq:d2J3dS2}
\end{align}
The term $\IItendev \sdev  +  \sdev \, \IItendev$ in \eqref{eq:d2J3dS2} reads in index notation $\IItendev_{imkl} s_{jm} + s_{im} \IItendev_{mjkl}$. The following second derivatives are needed to formulate the second derivative of the yield potential.
\begin{align}
\pddiff{A}{\sten} &= 
  \frac{\norm{\sten}^3\left(2 \sten \otimes \Iten 
- \Iten \otimes \sten 
- I_1 \IItensym\right) 
- \left(\norm{\sten}^2 \Iten 
- I_1 \sten\right) \otimes 3 \sten \norm{\sten}}{\sqrt{3}\norm{\sten}^6} \label{eq:d2AdS2} \\
\pddiff{B}{\sten} &=
  \frac{\pddiff{J_3}{\sten}}{J_2^{\nicefrac{3}{2}}}
- \frac{3}{2} \frac{\pdiff{J_3}{\sten} \otimes \pdiff{J_2}{\sten}}{J_2^{\nicefrac{5}{2}}}
- \frac{3}{2} \frac{\pdiff{J_2}{\sten} \otimes \pdiff{J_3}{\sten}}{J_2^{\nicefrac{5}{2}}}
- \frac{3}{2} J_3 \frac{\pddiff{J_2}{\sten}}{J_2^{\nicefrac{5}{2}}}
+ \frac{15}{4} J_3 \frac{\pdiff{J_2}{\sten} \otimes \pdiff{J_2}{\sten}}{J_2^{\nicefrac{7}{2}}}
\label{eq:d2BdS2} \\[1ex]
\pddiff{C}{\sten} &=
  \pdiff{A}{\sten} \otimes \pdiff{B}{\sten}
+ B \pddiff{A}{\sten}
+ \pdiff{B}{\sten} \otimes \pdiff{A}{\sten}
+ A \pddiff{B}{\sten}
\label{eq:d2CdS2} \\[1ex]
\pddiff{D}{\sten} &= 
\gamma \, m \left( 1 + \gamma C \right)^{m-2}
\left[
\left( 1 + \gamma C \right) \pddiff{C}{\sten} 
+ \gamma \left( m - 1 \right) \pdiff{C}{\sten} \otimes \pdiff{C}{\sten}
\right]
\label{eq:d2DdS2} \\[1ex]
\pddiff{E}{\sten} &= 
  \pdiff{D}{\sten} \otimes \pdiff{J_2}{\sten}
+ J_2 \pdiff{D}{\sten}
+ \pdiff{J_2}{\sten} \otimes \pdiff{D}{\sten}
+ D \pdiff{J_2}{\sten}
\label{eq:d2EdS2} \\[1ex]
\pddiff{W}{\sten} &= 
\frac{2 \left(f+g+h\right) \left(f''+g''+h''\right) - \left(f'+g'+h'\right) \otimes \left(f'+g'+h'\right)}{4 \left( f+g+h \right)^{\nicefrac{3}{2}}}
\label{eq:d2WdS2}
\end{align}
The symbols in \eqref{eq:d2WdS2} are placeholders for the following terms.
\begin{align}
f &= E, \quad f'=\pdiff{E}{\sten}, \quad f''= \pddiff{E}{\sten} \label{eq:fders} \\
g &= \frac{1}{2} \alpha I_1^2, \quad g'=\alpha I_1 \Iten, \quad g''=\alpha \Iten \otimes \Iten \label{eq:gders} \\
h &= \delta^2 I_1^4, \quad h'=4 \delta^2 I_1^3 \Iten, \quad h''=12 \delta^2 I_1^2 \Iten \otimes \Iten \label{eq:hders}
\end{align}
Finally, the second derivative of the yield potential with respect to the Cauchy stress tensor reads
\begin{align}
\pddiff{F}{\sten} &= \pddiff{W}{\sten} + 2 \epsilon \Iten \otimes \Iten\,.
\end{align}
In case that the parameters $\alpha$, $\beta$, $\gamma$, $\delta$, $\epsilon$, $\kappa$ and $m$ are functions of $\eqpl$ the derivative 
\begin{align}
\pdiff{F}{\eqpl} &= \pdiff{F}{\alpha} \pdiff{\alpha}{\eqpl} 
+ \pdiff{F}{\beta} \pdiff{\beta}{\eqpl} 
+ \pdiff{F}{\gamma} \pdiff{\gamma}{\eqpl} 
+ \pdiff{F}{\delta} \pdiff{\delta}{\eqpl} 
+ \pdiff{F}{\epsilon} \pdiff{\epsilon}{\eqpl} 
+ \pdiff{F}{\kappa} \pdiff{\kappa}{\eqpl}
+ \pdiff{F}{m} \pdiff{m}{\eqpl} 
\label{eq:dFdeqpl}
\end{align}
is needed, with
\begin{align}
\pdiff{F}{\alpha} &= \dfrac{I_1^2}{4\,W}\,, \label{eq:dFda}\\[1ex]
\pdiff{F}{\beta} &= I_1\,, \label{eq:dFdb}\\
\pdiff{F}{\gamma} &= \frac{C \, J_2 \, m \left( 1 + \gamma C \right)^{m-1}}{2 \, W}\,, \label{eq:dFdc}\\
\pdiff{F}{\delta} &= \frac{I_1^4 \, \delta}{W}\,, \label{eq:dFdd}\\[1ex]
\pdiff{F}{\epsilon} &= I_1^2\,, \label{eq:dFde}\\[1ex]
\pdiff{F}{\kappa} &= -1\,, \label{eq:dFdk}\\[1ex]
\pdiff{F}{m} &= \frac{J_2 \left( 1 + \gamma C \right)^m \log\left( 1+\gamma C \right)}{2 \, W}\,. \label{eq:dFdm}
\end{align}
In Eq.~\eqref{eq:LinReturnMappingMatrix3} the mixed derivatives of $F$ with respect to $\sten$ and $\eqpl$ are required. Here, we assume that all parameters depend on $\eqpl$.
\begin{align}
\pdddiff{F}{\sten}{\eqpl} &= \pdddiff{W}{\sten}{\eqpl} + \pdiff{\beta}{\eqpl} \Iten + 2 \pdiff{\epsilon}{\eqpl} I_1 \Iten
\label{eq:dFdSdeqpl}\\
\pdddiff{W}{\sten}{\eqpl} &= \frac{g \, f' - g' f}{2\,g^2}
\label{eq:dWdSdeqpl}
\end{align}
The symbols in Eq.~\eqref{eq:dWdSdeqpl} are placeholders for the following terms:
\begin{align}
f &= \pdiff{E}{\sten} + \alpha I_1 \Iten + 4 \delta^2 I_1^3 \Iten \\
f' &= \pdddiff{E}{\sten}{\eqpl} + \pdiff{\alpha}{\eqpl} I_1 \Iten + 8\,\delta \pdiff{\delta}{\eqpl} I_1^3 \Iten \\
g &= W \\
g' &= \pdiff{W}{\eqpl}
\end{align}
The mixed derivatives for $E$ and $D$ read as
\begin{align}
\pdddiff{E}{\sten}{\eqpl} &= J_2 \pdddiff{D}{\sten}{\eqpl} + \pdiff{D}{\eqpl} \pdiff{J_2}{\sten}\,, \label{eq:d2EdSdeqpl}\\[1ex]
\pdddiff{D}{\sten}{\eqpl} &= \pdddiff{D}{\sten}{\gamma} \pdiff{\gamma}{\eqpl} + \pdddiff{D}{\sten}{m} \pdiff{m}{\eqpl}\,, \label{eq:d2DdSdeqpl1}
\end{align}
with
\begin{align}
\pdddiff{D}{\sten}{\gamma} &= m \pdiff{C}{\sten} \left( 1 + \gamma \, C \right)^{m-2} \left( 1 + \gamma \, m \, C \right)\,, \label{eq:d2DdSdc}\\
\pdddiff{D}{\sten}{m} &= \gamma \pdiff{C}{\sten} \left( 1 + \gamma \, C \right)^{m-1} \left(m \, \log \left( 1 + \gamma \, C \right) + 1 \right)\,.  \label{eq:d2DdSdm}
\end{align}
In the above equations the derivatives of $W$, $E$, and $D$ with respect to $\eqpl$ are required and given as follows:
\begin{align}
\pdiff{W}{\eqpl} &= \dfrac{\pdiff{E}{\eqpl} + \dfrac{1}{2} \pdiff{\alpha}{\eqpl} I_1^2 + 2 \delta \pdiff{\delta}{\eqpl}}{2\,W}\,, \label{eq:dWdeqpl} \\[1ex]
\pdiff{E}{\eqpl} &= J_2 \pdiff{D}{\eqpl}\,, \label{eq:dEdeqpl} \\[1ex]
\pdiff{D}{\eqpl} &= \pdiff{D}{\gamma} \pdiff{\gamma}{\eqpl} + \pdiff{D}{m} \pdiff{m}{\eqpl}\,, \label{eq:dDdeqpl}
\end{align}
with
\begin{align}
\pdiff{D}{\gamma} &= m \, C \left( 1 + \gamma C \right)^{m-1}\,, \label{eq:dDdc}\\[1ex]
\pdiff{D}{m} &= \left( 1 + \gamma C \right)^m \log \left( 1 + \gamma C \right)\,. \label{eq:dDdm}
\end{align}
\subsubsection{Derivatives of the flow potential}
For a non-associative flow rule the data-driven approach sketched in section \ref{sec:PI} generates a flow potential in such a way that its derivatives with respect to the stress tensor, which is equivalent to the plastic flow direction, can be arbitrarily scaled. Therefore, it is considered that the flow direction is normalized using
\begin{align}
\Nten &= \left\| \pdiff{G}{\sten} \right\|^{-1} \pdiff{G}{\sten}\,.
\label{eq:normG}
\end{align}
The derivatives of the normalized flow direction $\Nten$ with respect to the stress tensor $\sten$ and the equivalent plastic strain $\eqpl$, which are necessary in Eq.~\eqref{eq:LinReturnMappingMatrix3} are then defined as
\begin{align}
\pdiff{\Nten}{\sten} &= \pdiff{G}{\sten} \otimes \pdiff{}{\sten} \left( \left\| \pdiff{G}{\sten} \right\|^{-1} \right) + \left\| \pdiff{G}{\sten} \right\|^{-1} \pddiff{G}{\sten}\,, \label{eq:dNdS} \\
\pdiff{\Nten}{\eqpl} &= \pdiff{G}{\sten} \pdiff{}{\eqpl} \left( \left\| \pdiff{G}{\sten} \right\|^{-1} \right) + \left\| \pdiff{G}{\sten} \right\|^{-1} \pdddiff{G}{\sten}{\eqpl}\,, \label{eq:dNdeqpl}
\end{align}
with
\begin{align}
\pdiff{}{\sten} \left( \left\| \pdiff{G}{\sten} \right\|^{-1} \right) &= \left\| \pdiff{G}{\sten} \right\|^{-3} \pddiff{G}{\sten} : \pdiff{G}{\sten}\,, \label{eq:dnNdS} \\
\pdiff{}{\eqpl} \left( \left\| \pdiff{G}{\sten} \right\|^{-1} \right) &= \left\| \pdiff{G}{\sten} \right\|^{-3} \pdddiff{G}{\sten}{\eqpl} : \pdiff{G}{\sten}\,. \label{eq:dnNdeqpl}
\end{align}
The parameters $\alpha$, $\beta$, $\gamma$, $\delta$, $\epsilon$, $m$, and $\kappa$ can be expressed by scalar functions $\hat{\alpha}(\eqpl)$, $\hat{\beta}(\eqpl)$, $\hat{\gamma}(\eqpl)$, $\hat{\delta}(\eqpl)$, $\hat{\epsilon}(\eqpl)$, $\hat{m}(\eqpl)$, and $\hat{\kappa}(\eqpl)$. These functions and their corresponding derivatives with respect to $\eqpl$ are specific for the porous structure to be modeled and need to be defined individually.
\subsection{Derivatives necessary for the parameter identification procedure}
\subsubsection{For the yield function}
For the parameter identification procedure the necessary derivatives for the Jacobian \eqref{eq:Jacobian} are
\begin{align}
\pdiff{F}{\pvec} &= \pdiff{F}{\alpha} + \pdiff{F}{\beta} + \pdiff{F}{\gamma} + \pdiff{F}{\delta} + \pdiff{F}{\epsilon} + \pdiff{F}{\kappa} + \pdiff{F}{m}
\label{eq:dFdP}
\end{align}
and
\begin{align}
\pdiff{F}{x} &= \pdiff{W}{x} + \beta \pdiff{I_1}{x} + \epsilon \pdiff{I_1^2}{x}\,,
\label{eq:dFdx}
\end{align}
with
\begin{align}
\pdiff{W}{x} &= \frac{\pdiff{E}{x} + \dfrac{1}{2} \alpha \pdiff{I_1^2}{x} + \delta^2 \pdiff{I_1^4}{x}}{2\,W}\,,\label{eq:dWdx}\\[1ex]
\pdiff{E}{x} &= \pdiff{J_2}{x} D\,,\label{eq:dEdx}
\end{align}
and
\begin{align}
\pdiff{I_1}{x} = I_1\,, \quad
\pdiff{I_1^2}{x} = 2 x I_1^2\,, \quad
\pdiff{I_1^4}{x} = 4 x^3 I_1^4\,, \quad 
\pdiff{J_2}{x} = 2 x J_2\,.
\label{eq:dInvdx}
\end{align}
For the Eqs.~\eqref{eq:dFdx} -- \eqref{eq:dInvdx} it is assumed that the usual stress argument $\sten$ is given by its scaled version $x \sten$.  
\subsubsection{For the plastic flow direction}
The derivative of the normalized plastic flow direction in \eqref{eq:JacobianG} is given as
\begin{align}
\pdiff{}{\pvec_{Gi}}\left( \dfrac{\Nten}{\norm{\Nten}} \right) = \dfrac{\pdiff{\Nten}{\pvec_{Gi}} \norm{\Nten}^2 - \Nten * \pdiff{\Nten}{\pvec_{Gi}} : \Nten}{\norm{\Nten}^3} \,,
\label{eq:dNGdP}
\end{align}
with
\begin{align}
\pdiff{\Nten}{\pvec_{Gi}} = \pdddiff{G}{\sten}{\pvec_{Gi}} = \left[ \pdddiff{G}{\sten}{\alpha_{Gi}},\,\pdddiff{G}{\sten}{\gamma_{Gi}},\,\pdddiff{G}{\sten}{m_{Gi}}\right]^T \,,
\label{eq:dNdP}
\end{align}
whereas
\begin{align}
\pdddiff{G}{\sten}{\alpha_{Gi}} &= \dfrac{I_1 \Iten \, G - \pdiff{G}{\alpha_{Gi}} \left( \pdiff{E}{\sten} + \alpha_{Gi} \, I_1 \Iten \right)}{2 \, G^2} \,, \label{eq:d2GdSda} \\
\pdddiff{G}{\sten}{\gamma_{Gi}} &= \dfrac{\pdddiff{E}{\sten}{\gamma_{Gi}} \, G - \pdiff{G}{\alpha_{Gi}} \left( \pdiff{E}{\sten} + \alpha_{Gi} \, I_1 \Iten \right)}{2 \, G^2} \,, \label{eq:d2GdSdc} \\
\pdddiff{G}{\sten}{m_{Gi}} &= \dfrac{\pdddiff{E}{\sten}{m_{Gi}} \, G - \pdiff{G}{\alpha_{Gi}} \left( \pdiff{E}{\sten} + \alpha_{Gi} \, I_1 \Iten \right)}{2 \, G^2} \,, \label{eq:d2GdSdm}
\end{align}
and
\begin{align}
\pdddiff{E}{\sten}{\gamma_{Gi}} &= J_2 \pdddiff{D}{\sten}{\gamma_{Gi}} + \pdiff{D}{\gamma_{Gi}} \pdiff{J_2}{\sten} \,, \label{eq:d2EdSdc}\\
\pdddiff{E}{\sten}{m_{Gi}} &= J_2 \pdddiff{D}{\sten}{m_{Gi}} + \pdiff{D}{m_{Gi}} \pdiff{J_2}{\sten} \,. \label{eq:d2EdSdm}
\end{align}
The derivatives $\pdddiff{D}{\sten}{\gamma_{Gi}}$ and $\pdddiff{D}{\sten}{m_{Gi}}$ are already defined in the Eqs.~\eqref{eq:d2DdSdc} and \eqref{eq:d2DdSdm}, whereas the parameters $\gamma$ and $m$ are to replaced by $\gamma_{Gi}$ and $m_{Gi}$.
Therewith remain
\begin{align}
\pdiff{G}{\alpha_{Gi}} = \dfrac{I_1^2}{4\,G} \,, \quad \pdiff{G}{\gamma_{Gi}} = \dfrac{1}{2\,G}\pdiff{E}{\gamma_{Gi}}\,, \quad \pdiff{G}{m_{Gi}} = \dfrac{1}{2\,G}\pdiff{E}{m_{Gi}} \,, \label{eq:dGdacm}
\end{align}
and
\begin{align}
\pdiff{E}{\gamma_{Gi}} = J_2 \pdiff{D}{\gamma_{Gi}}\,, \quad \pdiff{E}{m_{Gi}} = J_2 \pdiff{D}{m_{Gi}} \label{eq:dEdcm}\,,
\end{align}
using $\pdiff{D}{\gamma_{Gi}}$ and $\pdiff{D}{m_{Gi}}$ as defined in the Eqs.~ \eqref{eq:dDdc} and \eqref{eq:dDdm}.
\end{document}